\documentclass[epsfig,12pt]{article}
\usepackage{epsfig}
\usepackage{graphicx}
\usepackage{array}
\usepackage{color}
\usepackage[numbers,sort&compress]{natbib}
\usepackage{hyperref}
\hypersetup{
	colorlinks=true,  
	linktoc=page,
	linkcolor=blue
}
\usepackage[nottoc,notlof,notlot]{tocbibind} 
\usepackage[titles]{tocloft}

\usepackage{amssymb,amsmath,mathtools,physics,float,subcaption}
\allowdisplaybreaks

\def\d{\partial}
\def\CC{\mathbb{C}}
\def\RR{\mathbb{R}}

\def\cp{\mathbb{CP}}
\def\cK{\mathcal{K}}
\def\cL{\mathcal{L}}
\def\cM{\mathcal{M}}
\def\bphi{\overline{\phi}}

\def\bz{\overline{z}}
\def\Li#1{\operatorname{Li}_{#1}}

\def\vol{\operatorname{Vol}}
\def\arcsinh{\operatorname{arcsinh}}
\def\arccosh{\operatorname{arccosh}}
\def\sn#1#2{\operatorname{sn}^{#2}\left( #1 \right)}
\def\cn#1#2{\operatorname{cn}^{#2}\left( #1 \right)}
\def\dn#1#2{\operatorname{dn}^{#2}\left( #1 \right)}

\newcommand{\beq}{\begin{equation}}   
\newcommand{\eeq}{\end{equation}}
\newcommand{\beqn}{\begin{eqnarray}}   
\newcommand{\eeqn}{\end{eqnarray}}

\newcommand{\pt}{\partial}

\newcommand{\Rc}{{\mathcal R}}
\newcommand{\Lc}{{\mathcal L}}

\newcommand{\gsim}{\lower.7ex\hbox{$
\;\stackrel{\textstyle>}{\sim}\;$}}
\newcommand{\lsim}{\lower.7ex\hbox{$
\;\stackrel{\textstyle<}{\sim}\;$}}
\begin{document}

\begin{titlepage}

   \begin{flushright}
    FTPI-MINN-23-05, UMN-TH-4211/23\\
    \end{flushright}

\begin{center}
{\Large \bf Remarks on Baby Skyrmion
Lie-Algebraic  \\[2mm]Generalization }

\end{center}

 \vspace{5mm}
    
    \begin{center}
    { \bf   Chao-Hsiang Sheu$^{a}$ and Mikhail Shifman$^{a, b}$}
    \end {center}
    
    \begin{center}
    
        {\it  $^{a}$Department of Physics,
    University of Minnesota,
    Minneapolis, MN 55455}\\{\small and}\\
    {\it  $^{b}$William I. Fine Theoretical Physics Institute,
    University of Minnesota,
    Minneapolis, MN 55455}\\
    
    \end{center}

\vspace{10mm}

\begin{center}
{\bf Abstract}
\end{center} 

We discuss generalized baby Skyrmions emerging in a (1+2)-dimensional $\sigma$ model with a certain  Lie-algebraic structure. The same result applies to the Polyakov-Belavin instantons in $D=1+1$.  The O(3) symmetry of the target space is lost, but O(2) is preserved in the simplest model under consideration.  Both the topological charge and the soliton mass (the instanton action) are determined. Of special interest are  limiting cases of the deformation parameter $k$ (also referred to as the elongation parameter). If $|k-1|\ll1$, we arrive at a model which is studied in condensed matter. If $k\gg1$, we obtain the so-called cigar, or sausage model, well-known in little string theory. If $k=1$, we return to CP(1). The deformed model under consideration interpolates between CP(1) and the cigar model. In $D=2$ we calculate the coupling constants renormalization at one loop. At $k\geq 1$ this class of models is asymptotically free in the ultraviolet limit and enters the strong coupling domain in the infrared. Also, in the infrared $k\to 1$ (i.e., we recover CP(1)).

\end{titlepage}

\section{Introduction}
\label{intro}

Baby Skyrmions in $D=1+2$ attracted much attention in condensed matter physics, especially in magnetic phenomena.\footnote{The original baby Skyrmion is, in fact, the Polyakov-Belavin (PB) instanton \cite{BP} in the 2D CP(1) model elevated to $D=1+2$ dimensions and reinterpreted as a static soliton excitation.}  
The literature on this topic is enormous, including studies of the baby Skyrmion Hall effect of the texture and the topological Hall effect of the electron  (see, e.g., the review in \cite{GMT}). In this paper we consider the evolution of the baby Skyrmions as we deform the Heisenberg model (in the continuous limit) 
to include a certain Lie-algebraic  construction (see \cite{LMZ,Los}). Deformed in this way 
the baby Skyrmion continuously interpolates between the standard Polyakov-Belavin baby Skyrmion  and the so-called cigar models \cite{FOZ} (for a review see \cite{LZ}).

The standard Heisenberg model --a prototype model in this range of questions --can be written as\,\footnote{Also referred to as the O(3) model.}
\beq
{\mathcal L} =\frac{1}{2g^2} (\pt S_i) (\pt S_i)\,,\qquad \vec{S} \vec{S}=1\,,
\label{one}
\eeq
where $\vec S$ is a unit isovector describing spin, $\vec{S} =\{ S_1, S_2, S_3\}$,  and $g^2$ is the coupling constant with mass dimension $[m^{-1}]$. Its target space is $S^2$, a two-dimensional (2D) sphere.

Various generalizations of the standard  baby Skyrmions were considered previously (e.g., \cite{KH,BS} and references therein). Here we use the Lie-algebraic construction of \cite{Los,LMZ} to deform the O(3) model and, hence, the Polyakov-Belavin baby Skyrmion in a special way. Assuming that the deformed target space preserves O(2)$\times Z_2$ of the original O(3) we arrive at
\beq
{\mathcal L} =\frac{1}{2g^2(S_3)} (\pt S_i) (\pt S_i)\,,\qquad \vec{S} \vec{S}=1\,.
\label{two}
\eeq
where the coupling $g^2$ becomes a function of $S_3$, the third component of the isovector $\vec S$,
\beq
g^2(S_3) = g^2\cdot  \left(
\textstyle{\frac{1+k}{2}}+\textstyle{\frac{1-k}{2}}\,S_3^2\right) .
\label{three}
\eeq
Moreover,  $k$ is a numerical parameter to be defined below. At $k\to 1$ we return to the O(3) model while 
at $k\to+\infty$ with $g^2\cdot k =$const we approach the sausage model.
Equation (\ref{three}) leads to the following scalar curvature ${\mathcal  R}$ of the target space:
\begin{align}
  {\textstyle\frac{1}{ 2}} {\mathcal  R}
  &= g^2\,\frac{(1+k) - (1-k)S_3^2}{(1+k) + (1-k)S_3^2}
  \notag\\[2mm]
  &= g^2 \left[
  1+ \frac{2(k-1)S_3^2}{(1+k) + (1-k)S_3^2}\right].
\label{four}
\end{align}
When we consider $k\gg 1$, the scalar curvature is small in the equatorial region and almost everywhere else, $\sim g^2\sim 1/k$, with the exception of two polar domains
where $1-S_3^2\sim 1/k$. 
Indeed,
\beq
{\textstyle{\frac{1}{ 2}} {\mathcal  R}} \sim
 \left\{
\begin{array}{l} 
g^2 \,\,\,{\rm equator},
\\[1mm] g^2 k\,\,\,{\rm poles}.
\end{array}
\right.
\nonumber
\eeq
In the polar domains the scalar curvature $\sim g^2k\sim O(1)$, i.e., the ratio of the polar scalar curvature to equatorial, is $\sim k\gg 1$. 

Equations (\ref{two}) and (\ref{three}) do not present the most general extension. The extensions breaking O(2) of the target space [rotations around the third axis in the isospace in (\ref{two}), (\ref{three})]  can be obtained in a similar way; see footnote 3.

Organization of the paper is as follows. In Sec. \ref{lac} we review the Lie-algebraic construction that lies in the basis of the generalization under consideration. In Sec. \ref{bsi} and \ref{intstrbs}, we calculate the baby Skyrmion mass (equal to the instanton action in the two-dimensional model) and study the internal structure of the baby Skyrmions in detail. In Sec. \ref{lkl} we will study a special limit corresponding to the cigar models of \cite{FOZ}
also known as the metric of a 2D Euclidean black hole \cite{EFR,EW}.

\section{Lie-algebraic construction}
\label{lac}

The simplest Lie-algebraic construction is based on the $sl (2)\times sl(2)$ algebra
\cite{Los,LMZ}.
In the CP(1) model the target space is one-dimensional (two real dimensions) and is parametrized by a single complex field $\phi$ and its complex conjugated. Our task is to build a class of Lie-alegebaric extensions.
The  $sl(2)\times sl(2)$  generators can be represented in the form
\beqn
&& T^+=-\phi^2\,d_\phi\,,\quad T^0= \phi\,d_\phi\,,\quad T^-= d_\phi\,\nonumber\\[2mm]
&& \bar T^+=-\bar\phi^2\,d_{\bar\phi}\,,\quad \bar T^0= \bar\phi\,d_{\bar\phi}\,,\quad \bar T^-= d_{\bar\phi}\,.
\label{five}
\eeqn
where one of the two $sl(2)$ algebras is holomorphic and the other antiholomorphic.
The commutation relations are standard for $sl(2)$,
 \beq
[T^+,T^-] = 2 T^0\,,\quad [T^+,T^0] = - T^+\,,\quad [T^-,T^0] = + T^-\,,
\label{six}
\eeq 
with the structure constants 
\beq
f^{+-}_{\quad 0}=2, \quad f^{+0}_{\quad +}=-1, \quad  f^{-0}_{\quad -}=1,
\label{seven}
\eeq
and similar for $\bar T$s.

Generically, the Lie-algebraic metric (with the upper indices) must take the form of a quadratic in $T$  combination
\beq
G^{1\bar 1}  d_\phi d_{\bar\phi} = \sum_{a,\bar b} {\mathcal P}_{a\bar b} T^a \bar{T}^{\bar b}
\label{eight}
\eeq
with a set of numeric coefficients $\{ {\mathcal P}_{a\bar b} \}$.

 Assuming  that the target space preserves a residual U(1) symmetry,
 we can  reduce the set  of coefficients $\{ {\mathcal P}_{a\bar b} \}$ to the diagonal form
\beq
\{{\mathcal P}\} = \{
{\mathcal P}_{1\bar 1}\equiv n_1, \quad {\mathcal P}_{2\bar 2}\equiv n_2,\quad {\mathcal P}_{3\bar 3}\equiv n_3\}
\label{ten}
\eeq
with all  off-diagonal ${\mathcal P}_{a\bar b}$ vanishing. Without loss of generality we can choose the set of coefficients (\ref{ten}) real.
Then the metric with the lower indices $G_{1\bar 1}$ takes the form\,\footnote{In the most general case the Lie-algebraic metric $G_{1\bar 1}$  is parametrized as
$$
G_{1\bar 1}= \frac{1}{n_1+n_2 \bar\phi \phi +n_3 \bar\phi^2\phi^2
+\left(m_1\phi+ m_2\phi^2+m_3 \phi\bar\phi^2 +
{\rm H.c.} \right) }
$$}
\beq
G_{1\bar 1} = \frac{1}{n_1 +n_2\bar\phi \phi + n_3 (\bar\phi\phi)^2}\,\,.
\label{1one}
\eeq
We arrive at the Lagrangian
\beq
{\mathcal L} =G_{1\bar 1}\left(\pt_\mu\bar\phi \pt^\mu\phi\right).
\label{1two}
\eeq
If the coefficients $n_{1,3}$ are nonsingular (i.e. neither 0 nor $\infty$), by rescaling the fields $\phi,\bar\phi$,
\beq
\phi, \bar\phi\to \lambda \phi, \lambda\bar\phi,\qquad \lambda^2 =\sqrt{\frac{n_1}{n_3}}
\eeq
one can always make the first and the third coefficients equal to each other.\footnote{This may not be the case if, say, $n_1\to 0 $; see below.}  One can keep this in mind.
If so, one  can conveniently  parametrize $n_{1,2,3}$ as follows:
 \beq
n_1= n_3= \frac{g^2}{2}, \quad  n_2 = g^2k\,.
\label{twop}
 \eeq
 The only extra parameter compared to CP(1) is $k$. If $k =1$, we return to CP(1). Another interesting limit to be discussed below is $k\to +\infty$.

The geometry of the space (\ref{1one}) is K\"ahlerian. The K\"ahler potential is
  \begin{multline}
     \cK = -\frac{1}{g^2\sqrt{k^2-1}}\Bigg\{ 
         \Bigg[\log(-\frac{\phi\bphi}{\sqrt{k^2-1}+k})
         \log(\phi\bphi+\sqrt{k^2-1}+k)\\[1mm]
         +\Li{2}\left( \frac{\phi\bphi+\sqrt{k^2-1}+k}{\sqrt{k^2-1}+k} \right)\Bigg]
         - \left( k +\sqrt{k^2-1}\to k -\sqrt{k^2-1}\, \right)
     \Bigg\}
     \label{eq:kahler}
 \end{multline}
for $k\in [1,\infty]$. Here $\Li{2}(z)$ is the dilogarithm.
Further geometric data are given by the following expressions:
\begin{align}
\label{eq:curvature}
    &\Gamma^{1}_{11} = -\frac{2(k+\bphi\phi)\bphi}{1+2k \bphi\phi+3(\bphi\phi)^2}\,,
    \notag
    &&R_{1\bar{1}1\bar{1}} =\! -\frac{4\left[k + 2\phi\bphi+k(\phi\bphi)^2\right]}{g^2\left[(\bphi\phi)^2 + 2k \bphi\phi + \right]^3} \,,\quad\quad\notag\\[2mm]
    &R_{1\bar{1}}\, = \,\frac{2\left[k + 2\phi\bphi+k(\phi\bphi)^2\right]}{\left[(\bphi\phi)^2 + 2k \bphi\phi + 1\right]^2} \,,
    \quad
    &&{\mathcal R}\, = \,\frac{2g^2\left[k + 2\phi\bphi+k(\phi\bphi)^2\right]}{(\bphi\phi)^2 + 2k \bphi\phi + 1}\,.
\end{align}
Correspondence between $\phi, \bar\phi$  and the O(3) representation through the  unit vector ${\vec S = \{S_i\}}$, $i=1,2,3$   (see Eq. (\ref{one})) is realized through the stereographic projection,
\beq
\phi= \frac{S_1+ iS_2}{1+S_3}\,,\quad \bar\phi= \frac{S_1- iS_2}{1+S_3}\,.
\label{17}
\eeq
Then the following equations ensue:
\beqn
\pt_\mu\bar \phi\,\pt^\mu \phi= 
\frac{1}{\left(1+S_3\right)^{2}}\, 
\pt_\mu\vec{S}\,\pt^\mu \vec{S}\,,
\label{18}
\eeqn
and 
\beqn
&&n_1+n_2(\bar\phi\phi ) + n_3(\bar\phi\phi )^2 
= g^2 \frac{1}{\left(1+S_3\right)^2}\left[(1+k) + (1-k) S_3^2
\right].
\label{19}
\eeqn
Combining (\ref{18}) and (\ref{19}) we arrive at Eqs. (\ref{two}) and (\ref{three}). 
For future convenience, let us also present the spherical coordinate representation of the spin vector. Namely,
\begin{align}\label{ssph}
  S_1 = \sin{\alpha}\cos{\beta} \,,~
  S_2 = \sin{\alpha}\sin{\beta} \,,~
  S_3 = \cos{\alpha}
\end{align}
where $\alpha$ is the angle from the $z$ axis and $\beta$ is the azimuthal angle.

\section{Baby Skyrmions (Instantons in 2D)}
\label{bsi}

In the subsequent discussion, we will focus on $k \geq 1 $. 

The topological charge $Q$ of the 
baby Skyrmion [which is the same as that of 2D deformed $\cp(1)$] is defined by the pullback of the K\"ahler form on the target space (with suitable normalization) \cite{Perelomov:1987va}, say,
\begin{align}\label{20}
  Q &= \frac{1}{\vol(\cM)}\int_{\cM}\frac{\dd^{2}\phi}{(\bphi\phi)^2 + 2k \bphi\phi + 1}
  \notag\\[2mm]
  &= \frac{1}{\vol(\cM)}\int_{S^2}\frac{\abs{\pdv*{\phi}{z}}^2-\abs{\pdv*{\phi}{\bz}}^2}{(\bphi\phi)^2 + 2k \bphi\phi + 1} \, \dd^{2}{z}\,.
\end{align}
Here $\vol(\cM)$ stands for  the volume of the target space $\cM$,
\begin{align}\label{21}
    \vol(\cM) = \frac{2\pi \arccosh{k}}{\sqrt{k^2-1}}
\end{align}
Two spatial dimensions ($x$ and $y$)  are parametrized by complex variables $z,\bz$. 
 The duality equation is the same as in the Polyakov-Belavin analysis \cite{BP}, and so is the instanton solution; see below Eq. \eqref{23}.
Then the corresponding instanton action reads 
\begin{align}
  S_{\rm inst} = \frac{2}{g^2}\int_{\cM}\frac{\d_{\mu}\phi \, \d^{\mu}\bphi{}}{(\bphi\phi)^2 + 2k \bphi\phi + 1}
  \dd^{2}z
  = \frac{2\vol(\cM)}{g^2} \,.
\end{align}
In 1+2 dimensions the 2D instanton action is reinterpreted as the baby Skyrmion mass $M_{\rm baby \, Sk}$; since the baby Skyrmion
saturates the BPS bound we have
\begin{align}
  M_{\rm baby \, Sk} =
    \displaystyle\frac{4\pi}{g^2}\frac{\arccosh{k}}{\sqrt{k^2-1}} \,,
  \quad k \geq 1\,,
  \label{22}
\end{align}
see Fig. \ref{fig:scalarrn2}.

Alternatively, the same result could be obtained directly, with no reference to the topological charge and the BPS saturation. 
Indeed, the duality equation remains the same as in the CP(1) model implying that $\phi$ is an analytic function of $z$ (the sum of poles) and 
$\bar\phi$ is the complex conjugated analytic function of $\bar z$. The minimal soliton presents just a single pole.
The appropriate solution can be chosen as follows:
\begin{figure}[t] 
    \centering 
    \includegraphics[width=8cm]{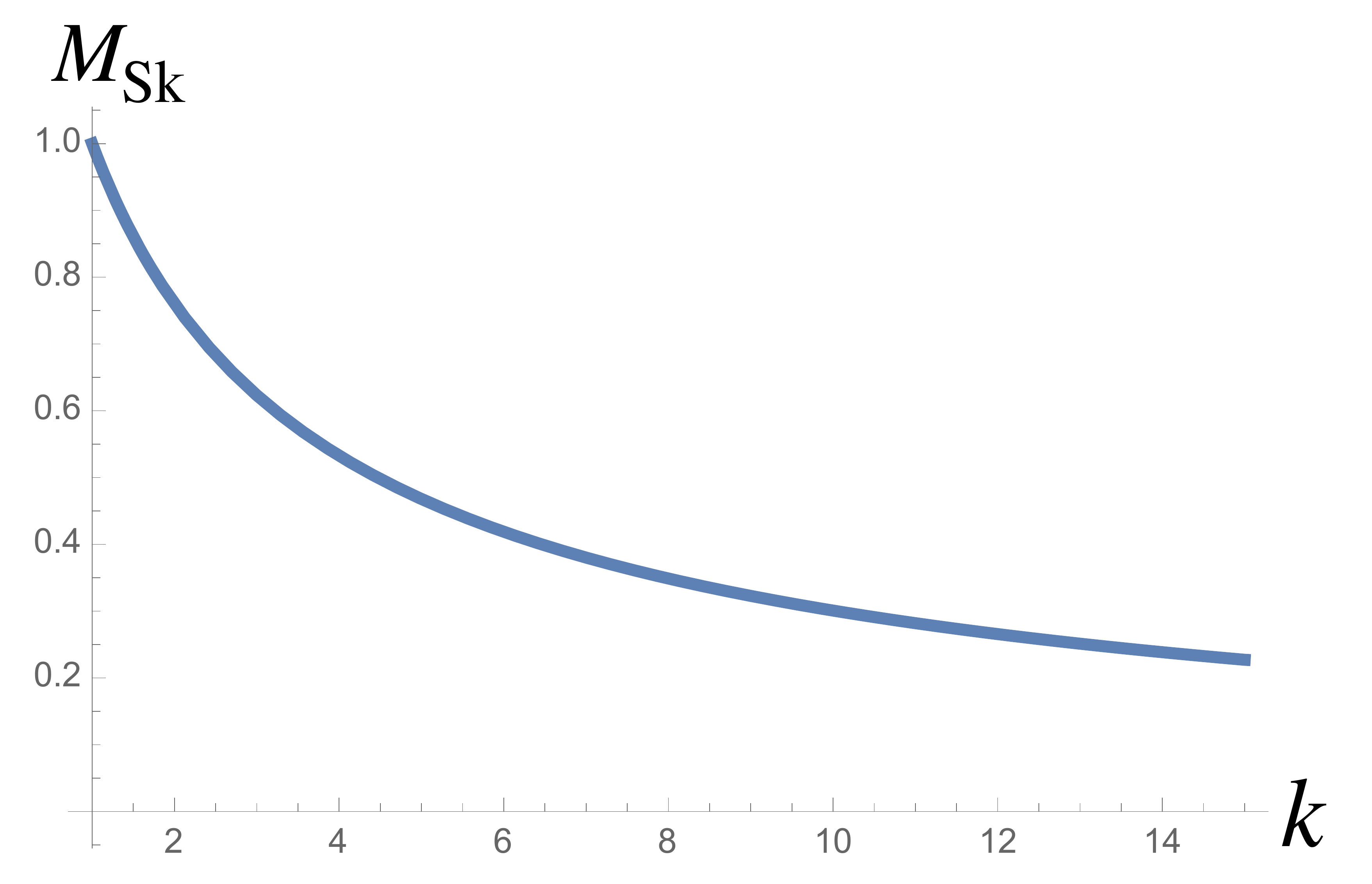}
    \caption{\small The baby Skyrmion mass in the units of $4\pi/g^2$. If $k\to\infty$ the Skyrmion mass tends to zero as $k^{-1}\log 2k$.  }
    \label{fig:scalarrn2}
  \end{figure}
\beq
\phi(z) = \frac{a}{z-z_0} 
\label{23}
\eeq
The parameter $|a|$ has the meaning of the overall size of the solution, and the phase of this parameter arg($a$) is a collective coordinate
reflecting the U(1) symmetry of the target space. We have four collective coordinates overall. After plugging the ansatz  \eqref{23} (with $z_0$ set to zero) in Eqs. (\ref{two}) and (\ref{three}), we arrive at the following  the density:
\begin{align}
    d\rho_{2}(x,y) 
    = d\tilde{z} d\bar{\tilde{z}}\,\, \frac{2}{g^2} \,\frac{1}{1 + 2k \,|{\tilde z}^2|+ |{\tilde z}^2|^2}
    \,,
    \label{24}
\end{align}
where
\beq
\tilde z = \frac{z}{a}\,.
 \label{25}
\eeq
After integrating the density above over the $\{x,y\}$ plane, we obtain the same mass as in \eqref{22}. 
The plots in Fig. \ref{fig:babysk} show the density distribution for different $k$.

\begin{figure*}[t] 
  \centering 
  \begin{subfigure}[b]{0.3\textwidth}
    \centering 
    \includegraphics[width=.9\textwidth]{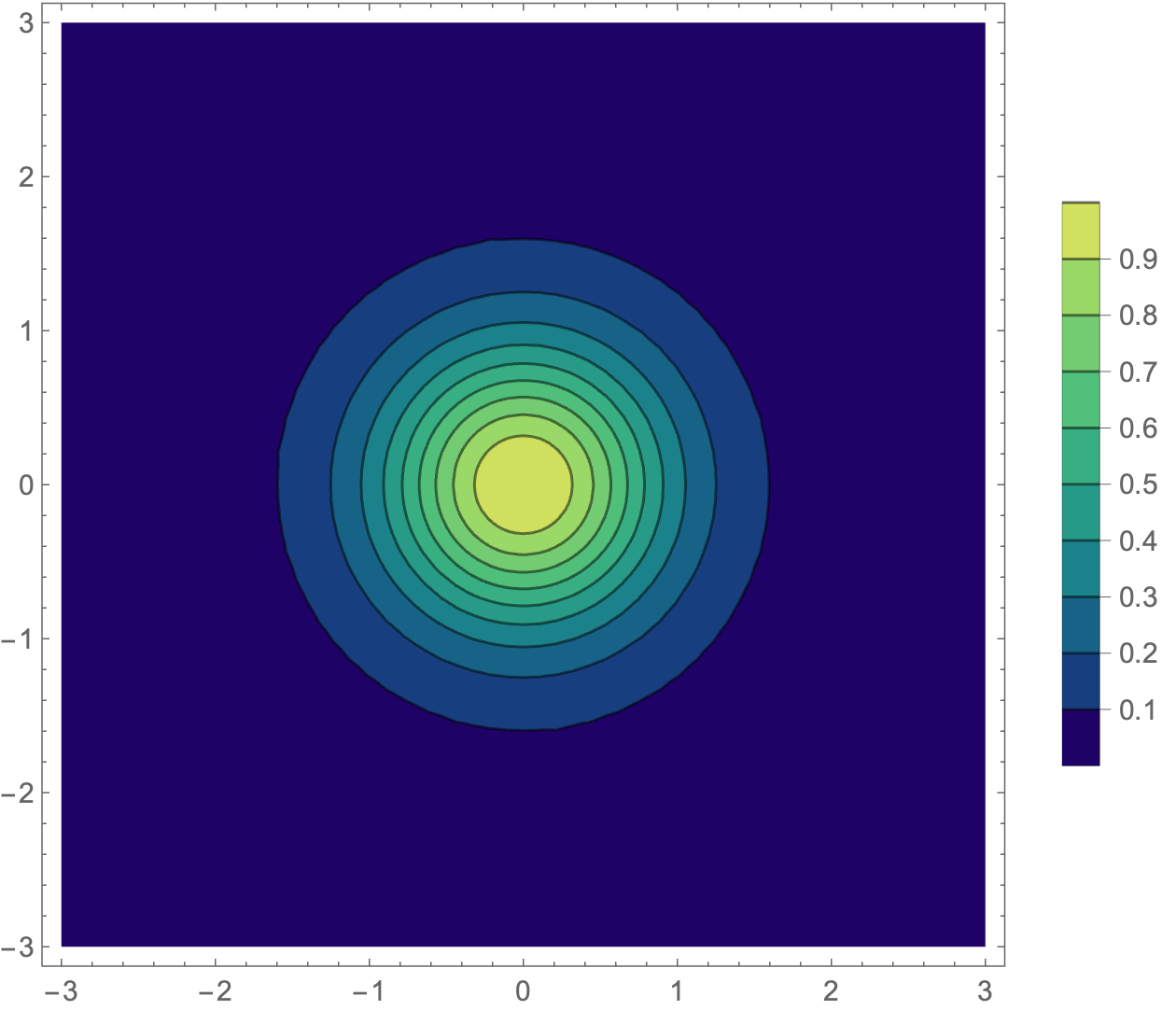}
    \caption{$k=0.5$}
  \end{subfigure}
  \begin{subfigure}[b]{0.3\textwidth}
    \centering 
    \includegraphics[width=.9\textwidth]{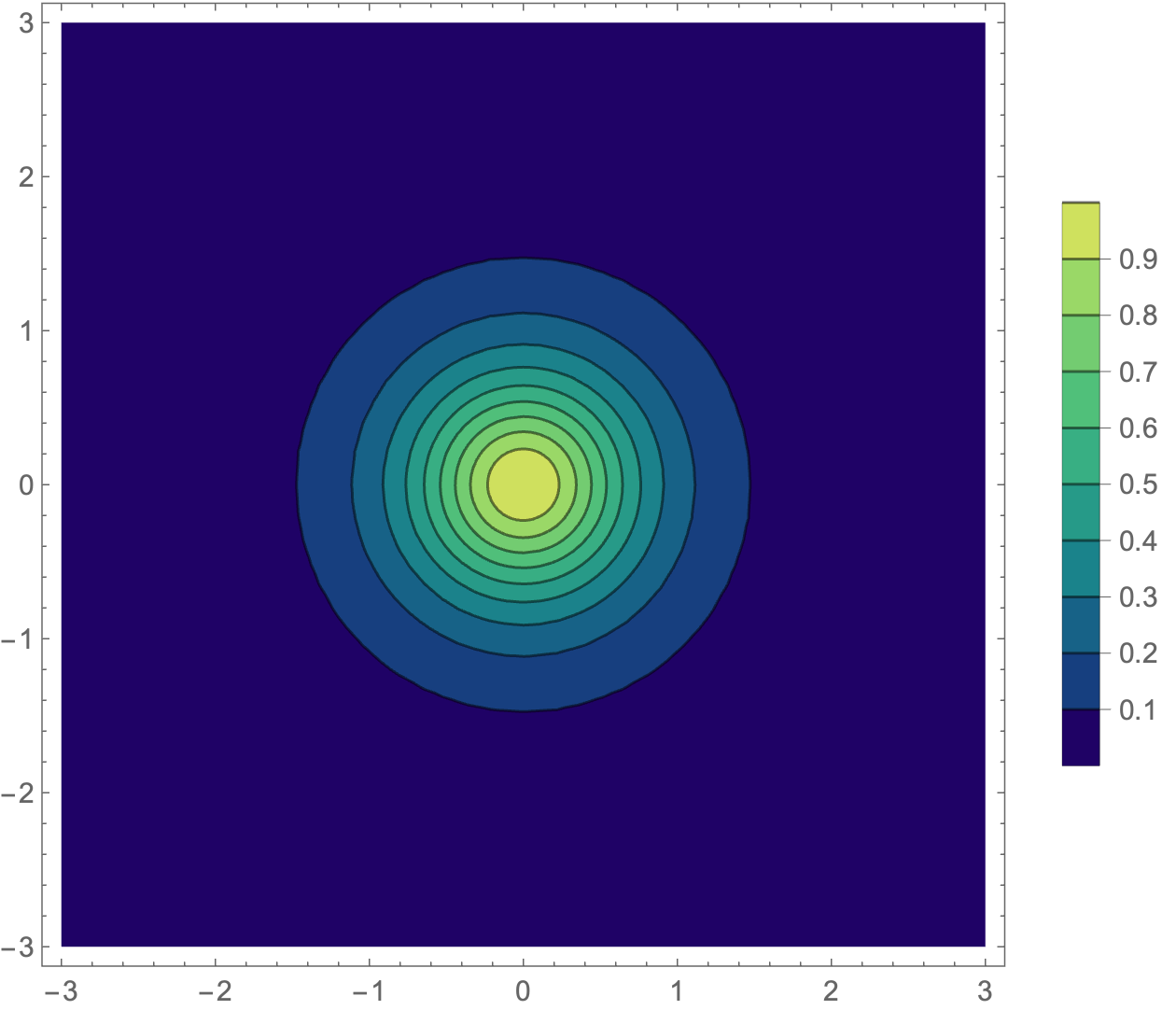}
    \caption{$k=1$}
  \end{subfigure}
  \begin{subfigure}[b]{0.3\textwidth}
    \centering 
    \includegraphics[width=.9\textwidth]{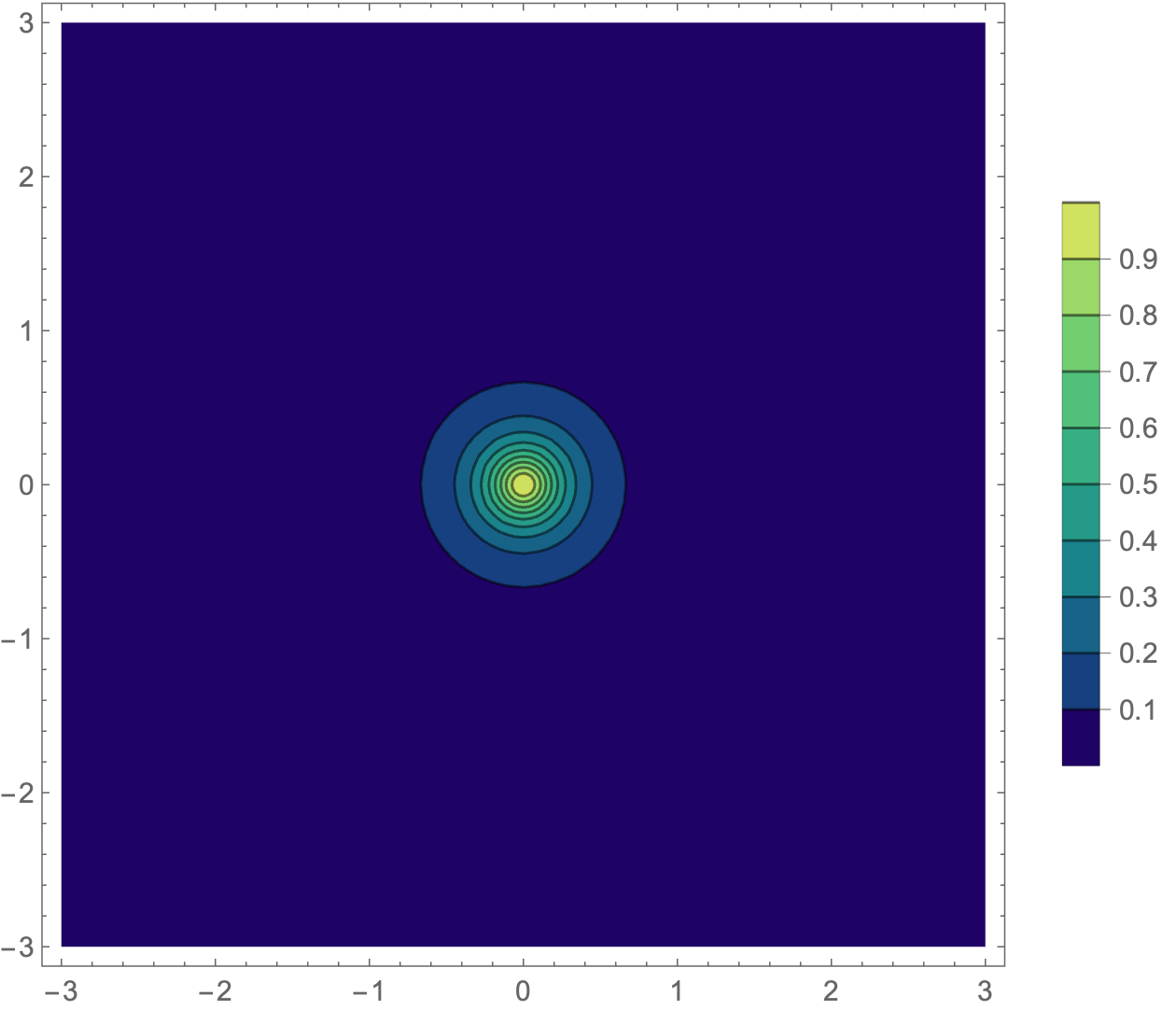}
    \caption{$k=2$}
  \end{subfigure}
  \caption{\small The density distribution of the Skyrmion solution for $k=0.5,1,2$. $a$ and $z_0$ are set at $1$ and $0$, respectively.}
  \label{fig:babysk}
\end{figure*}

\section{Internal structure of baby Skyrmions}
\label{intstrbs}

In this section, we further investigate the baby Skyrmion solutions discovered in the previous section. As shown in the metric \eqref{1one} of the deformed model, we can see that the vacuum moduli space of baby Skyrmion solutions, i.e., the target space, is a squashed sphere (see also Fig. \ref{sausage}). In contrast to the CP(1) case, where the moduli space is a round sphere and any chosen base point yields the same baby Skyrmion solution, the vacuum moduli space of the present model exhibits a different behavior. Varying selections of a reference point on its moduli space generate different baby Skyrmion configurations.

To gain insight into the above claim, let us examine the general solution of the unit charge baby Skyrmion, as presented in \cite{Eto:2009bz}. Namely, 
\begin{align}\label{c1soln}
  \phi = \frac{a}{z-z_0} + b
\end{align}
where $a,z_0,b \in \CC$. A choice of parameters, $a,z_0$, and $b$, corresponds to the size, position, and the orientation of the spin vector $\vec{S}$ in asymptotics $z \to \infty$, respectively.
Indeed, recall that via the stereographic map \eqref{17}, we know that the moduli field $\phi$ can be expressed in terms of the spin vector.
Comparing \eqref{c1soln} with \eqref{17}, in the infinite $z$ limit ($z_0$  is kept finite) we arrive at
\begin{align}
  \label{svev}
    \frac{S_{1,{\rm vev}} + i S_{2,{\rm vev}}}{1 - S_{3,{\rm vev}}}
    =
    b
\end{align}
indicating that the vacuum expectation value of $\vec{S}$ is determined by $b$.

\begin{figure*}[t] 
  \centering 
  \begin{subfigure}[b]{0.19\textwidth}
    \centering 
    \includegraphics[width=\textwidth]{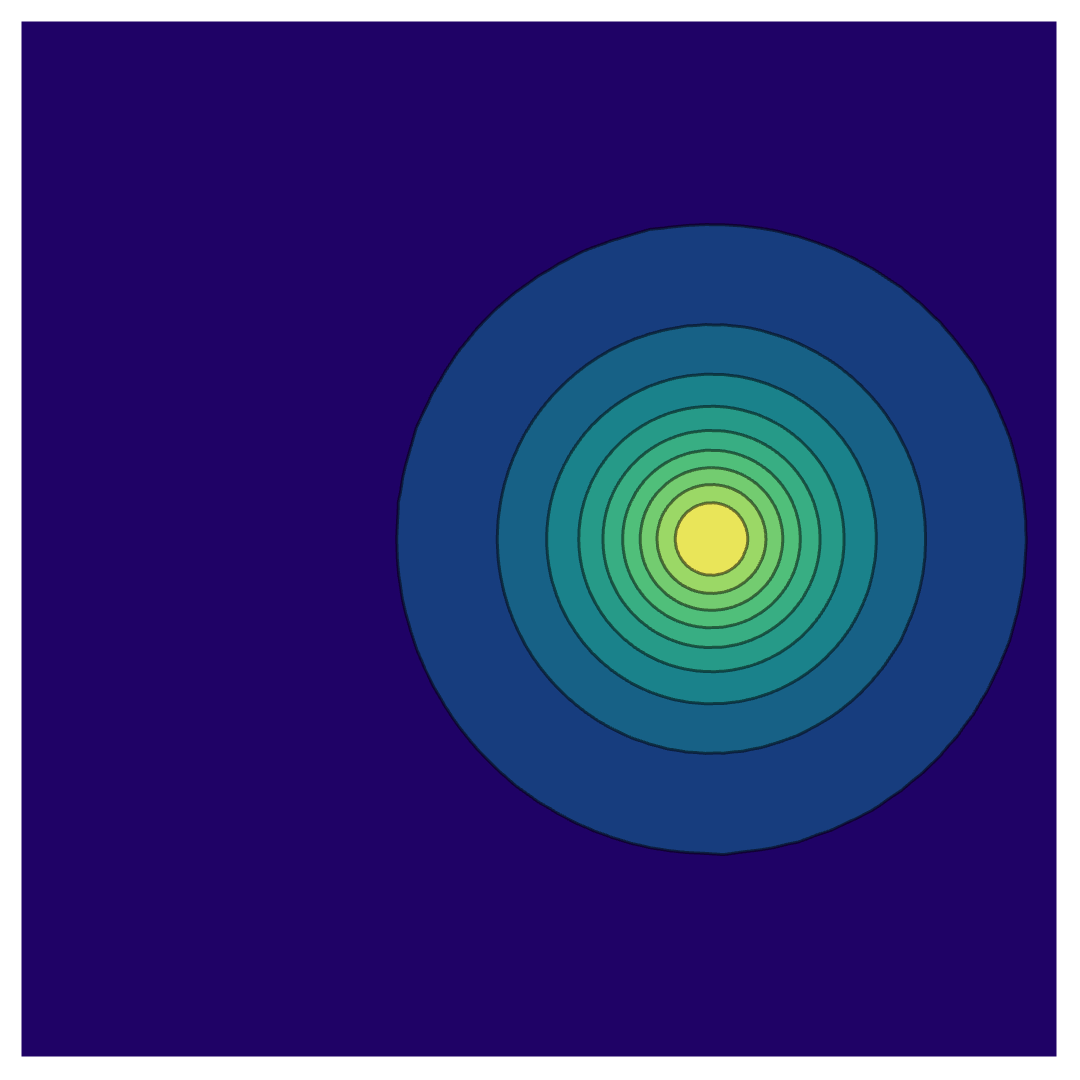}
  \end{subfigure}
  \begin{subfigure}[b]{0.19\textwidth}
    \centering 
    \includegraphics[width=\textwidth]{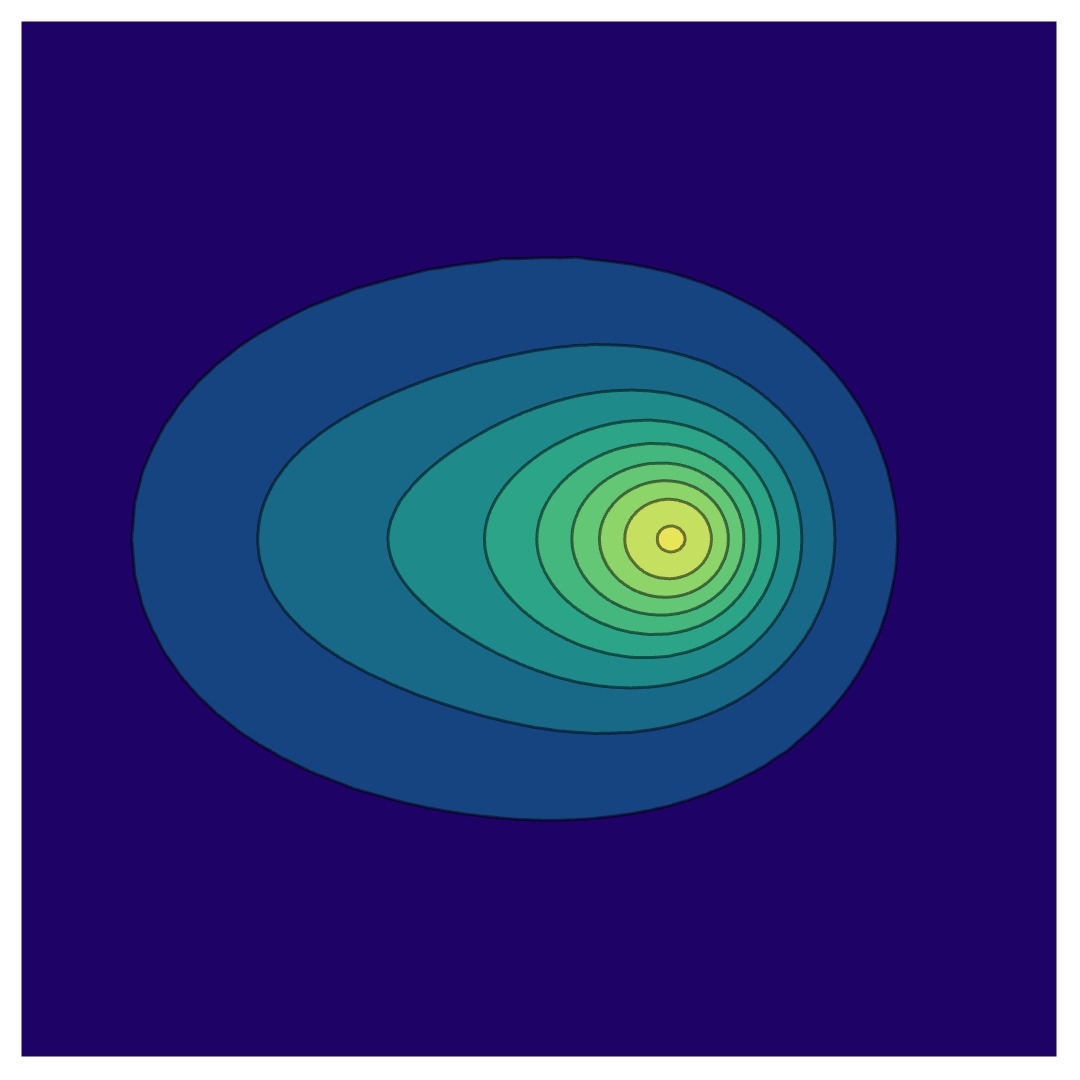}
  \end{subfigure}
  \begin{subfigure}[b]{0.19\textwidth}
    \centering 
    \includegraphics[width=\textwidth]{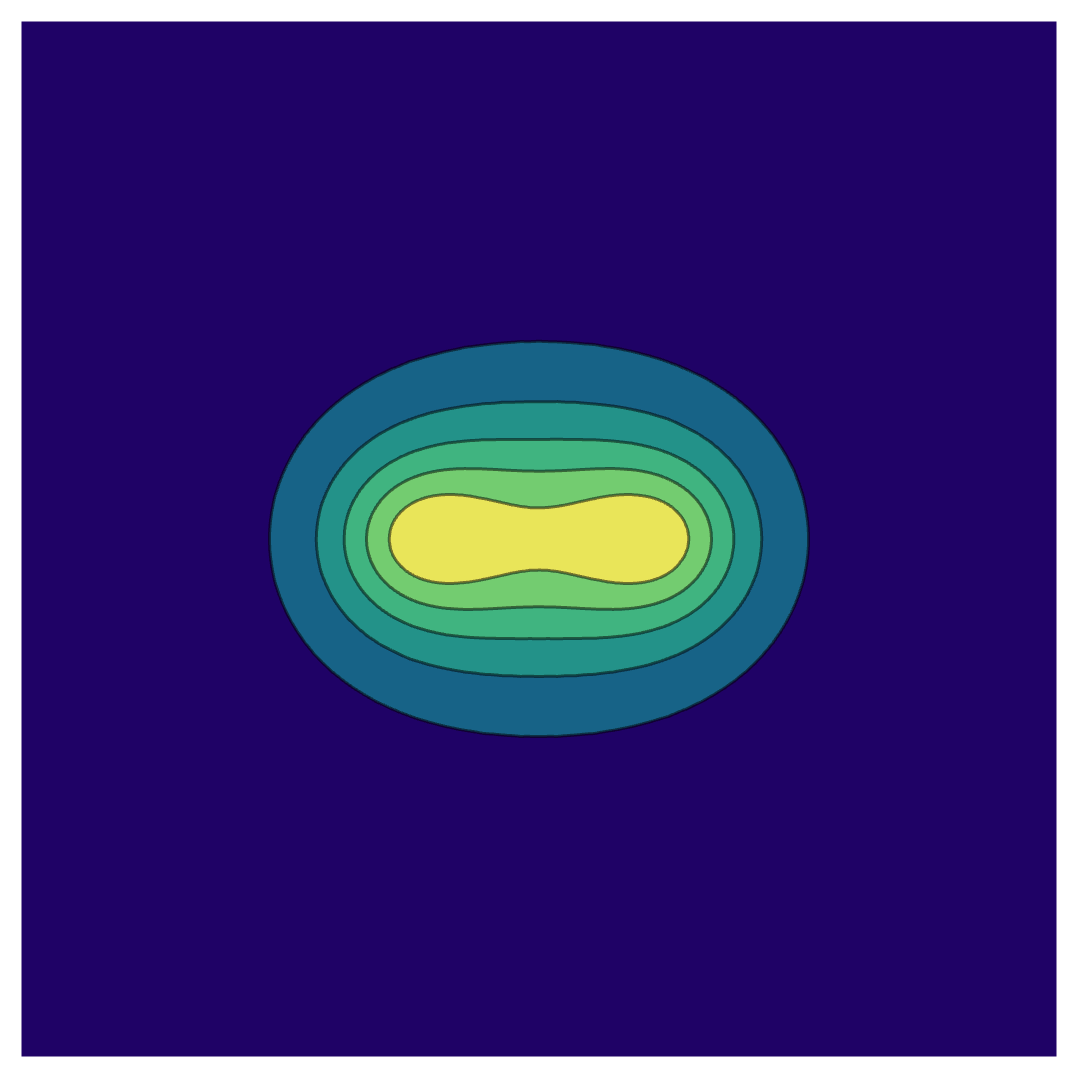}
  \end{subfigure}
  \begin{subfigure}[b]{0.19\textwidth}
    \centering 
    \includegraphics[width=\textwidth]{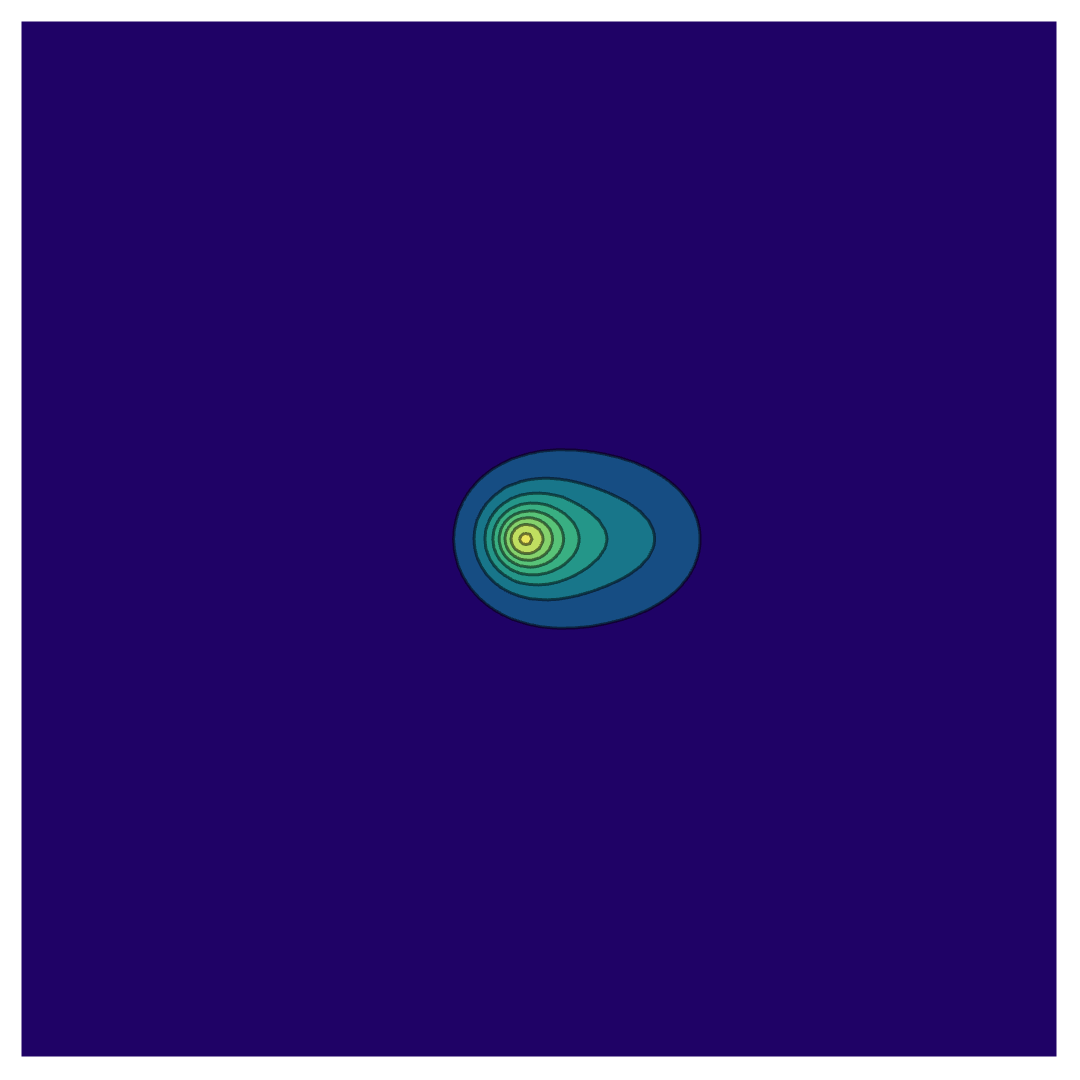}
  \end{subfigure}
  \begin{subfigure}[b]{0.19\textwidth}
    \centering 
    \includegraphics[width=\textwidth]{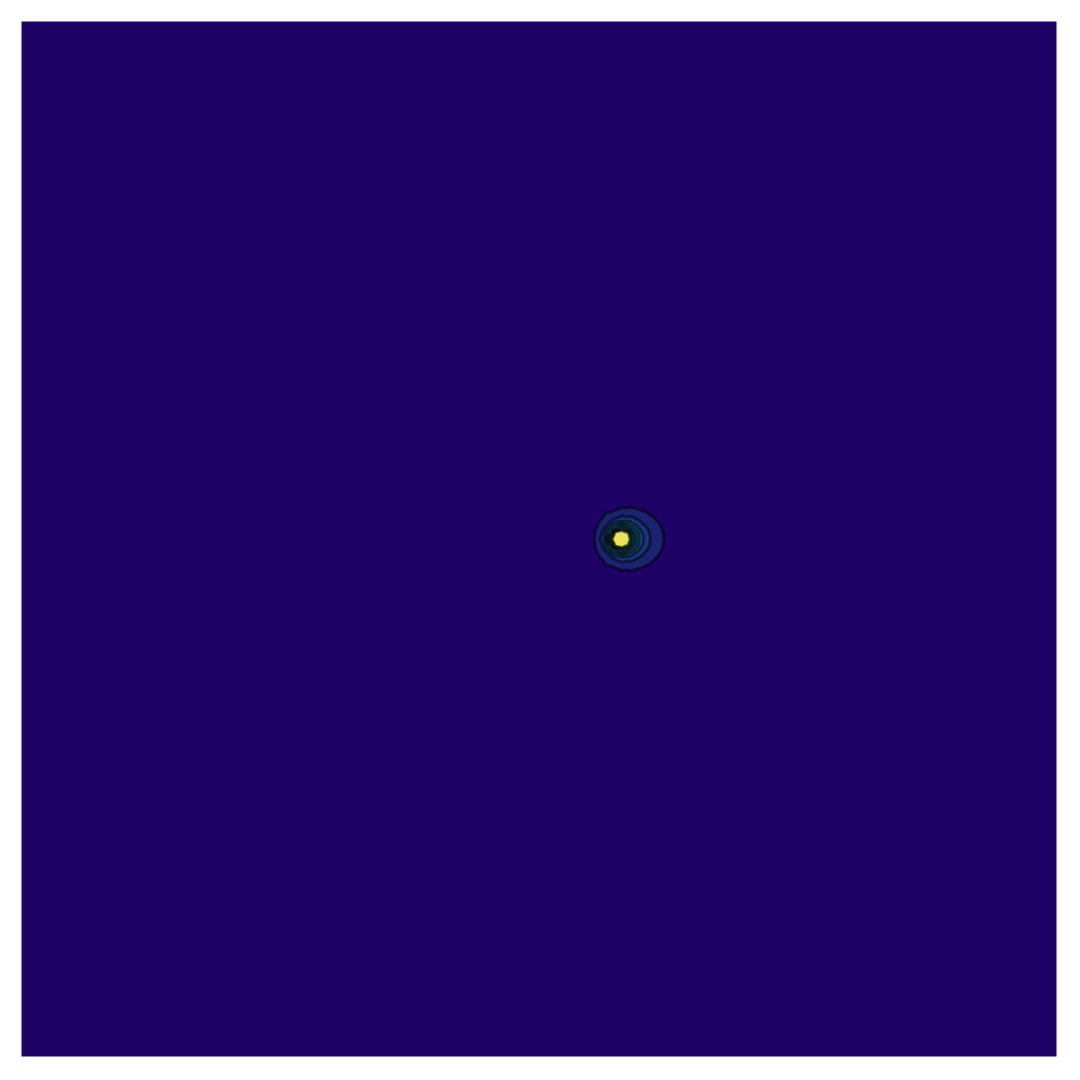}
  \end{subfigure}
  \\
  \begin{subfigure}[b]{0.19\textwidth}
    \centering 
    \includegraphics[width=\textwidth]{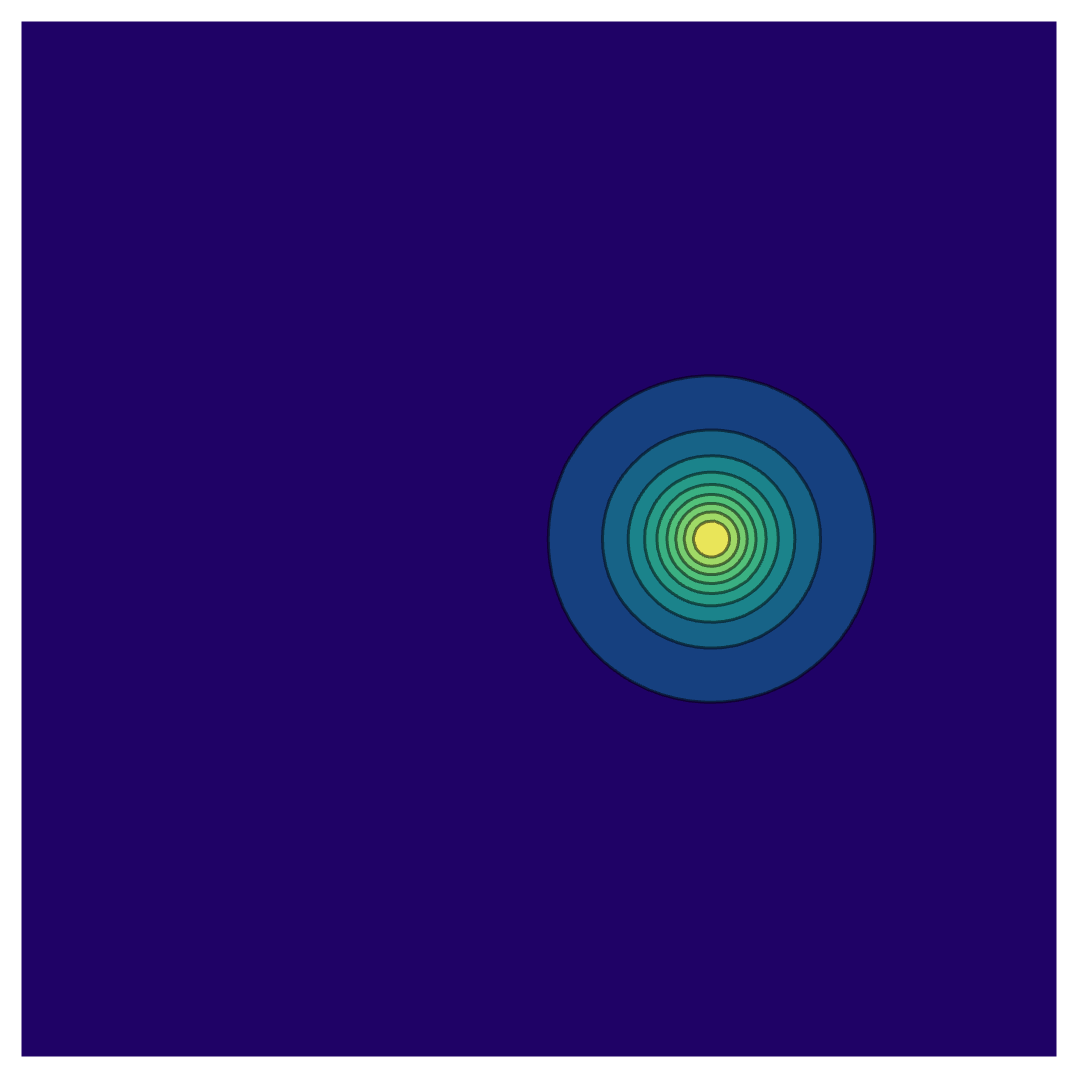}
  \end{subfigure}
  \begin{subfigure}[b]{0.19\textwidth}
    \centering 
    \includegraphics[width=\textwidth]{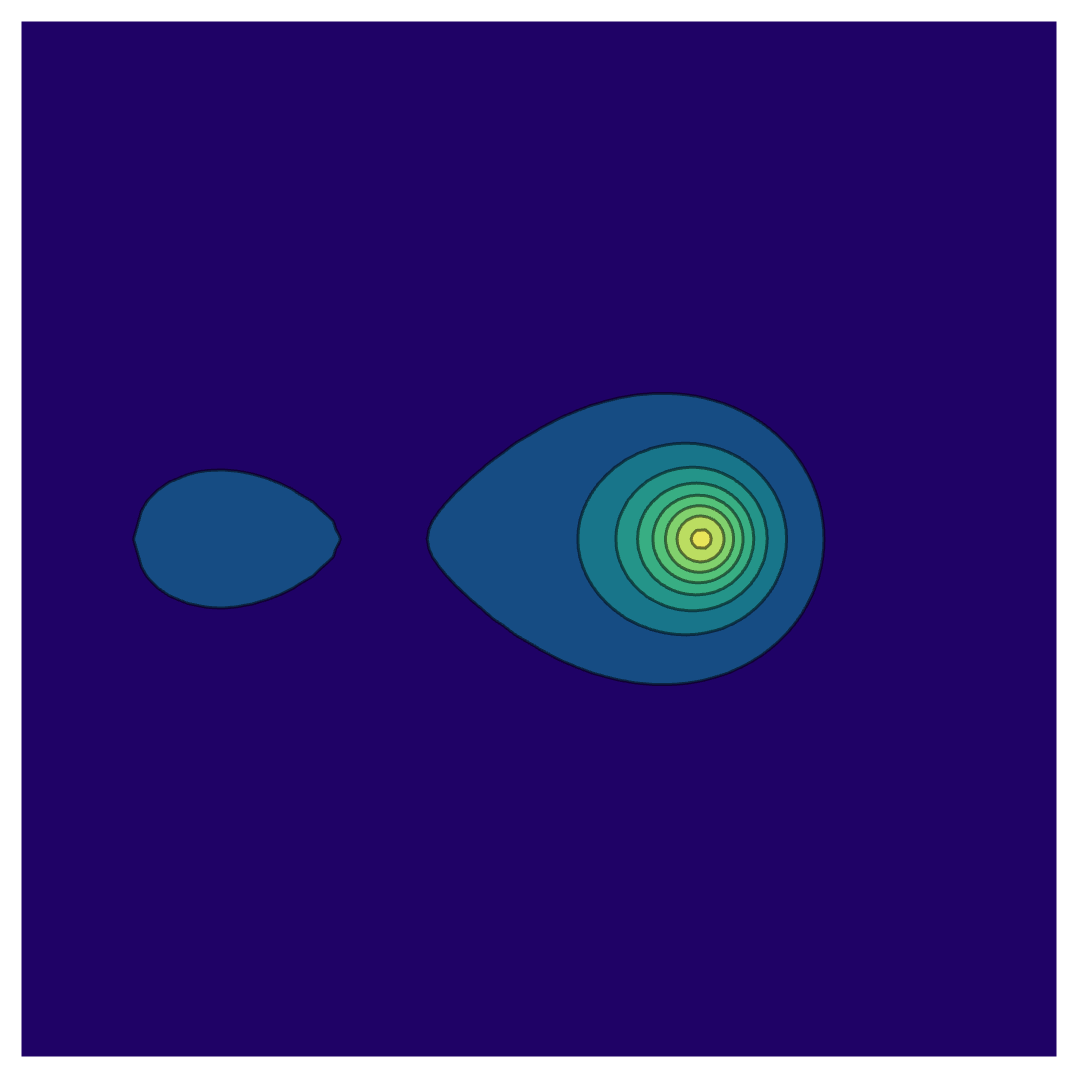}
  \end{subfigure}
  \begin{subfigure}[b]{0.19\textwidth}
    \centering 
    \includegraphics[width=\textwidth]{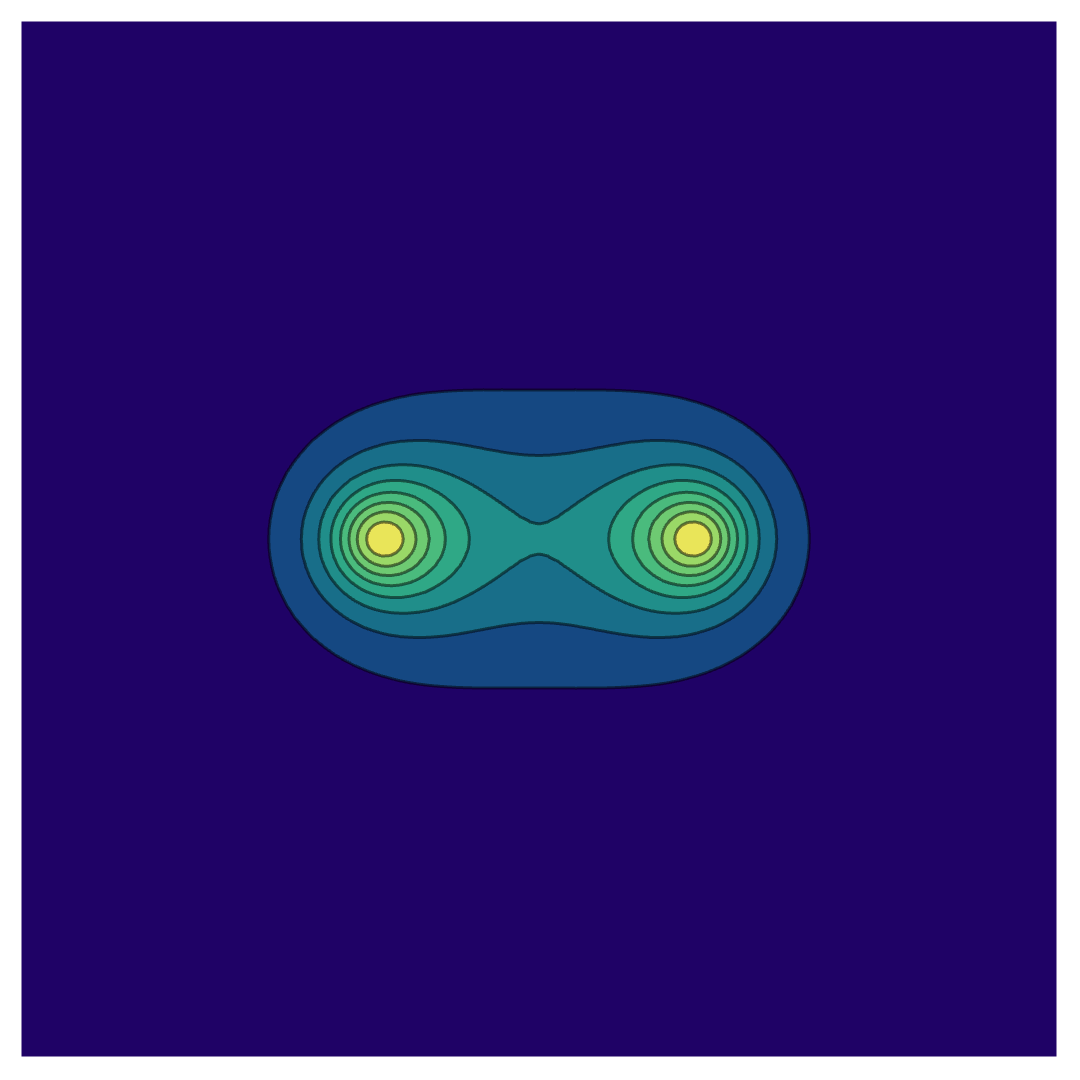}
  \end{subfigure}
  \begin{subfigure}[b]{0.19\textwidth}
    \centering 
    \includegraphics[width=\textwidth]{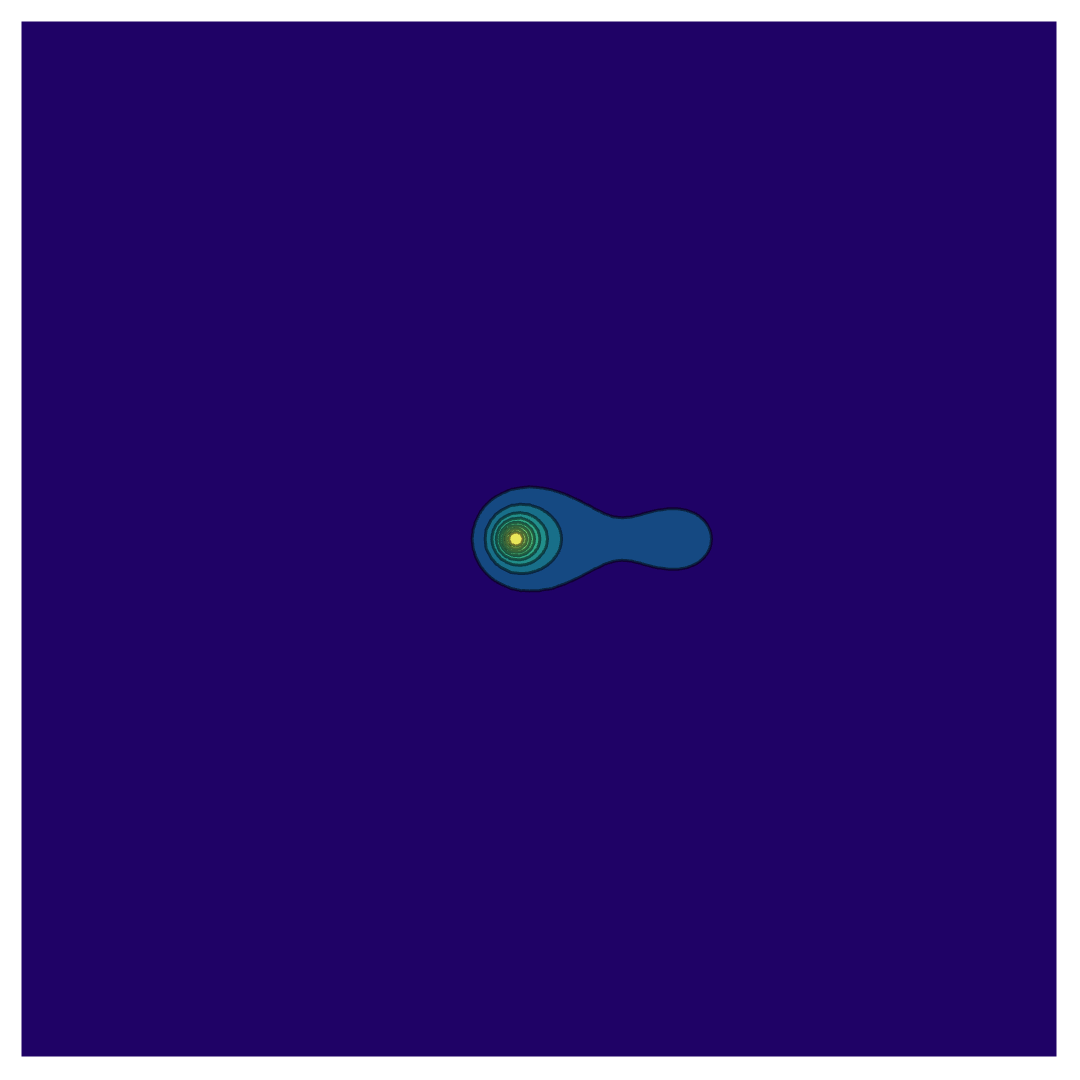}
  \end{subfigure}
  \begin{subfigure}[b]{0.19\textwidth}
    \centering 
    \includegraphics[width=\textwidth]{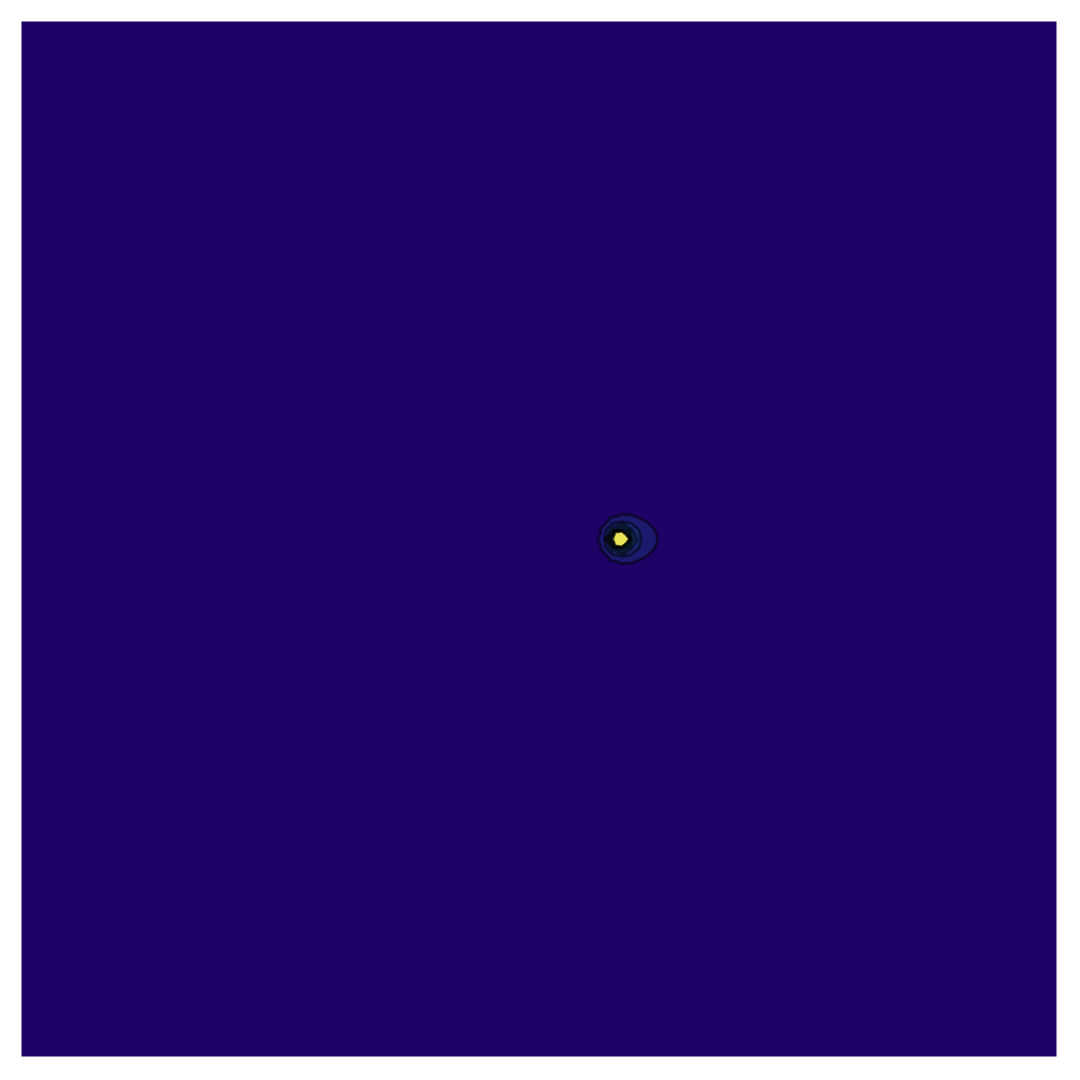}
  \end{subfigure}
  \\ 
  \begin{subfigure}[b]{0.19\textwidth}
    \centering 
    \includegraphics[width=\textwidth]{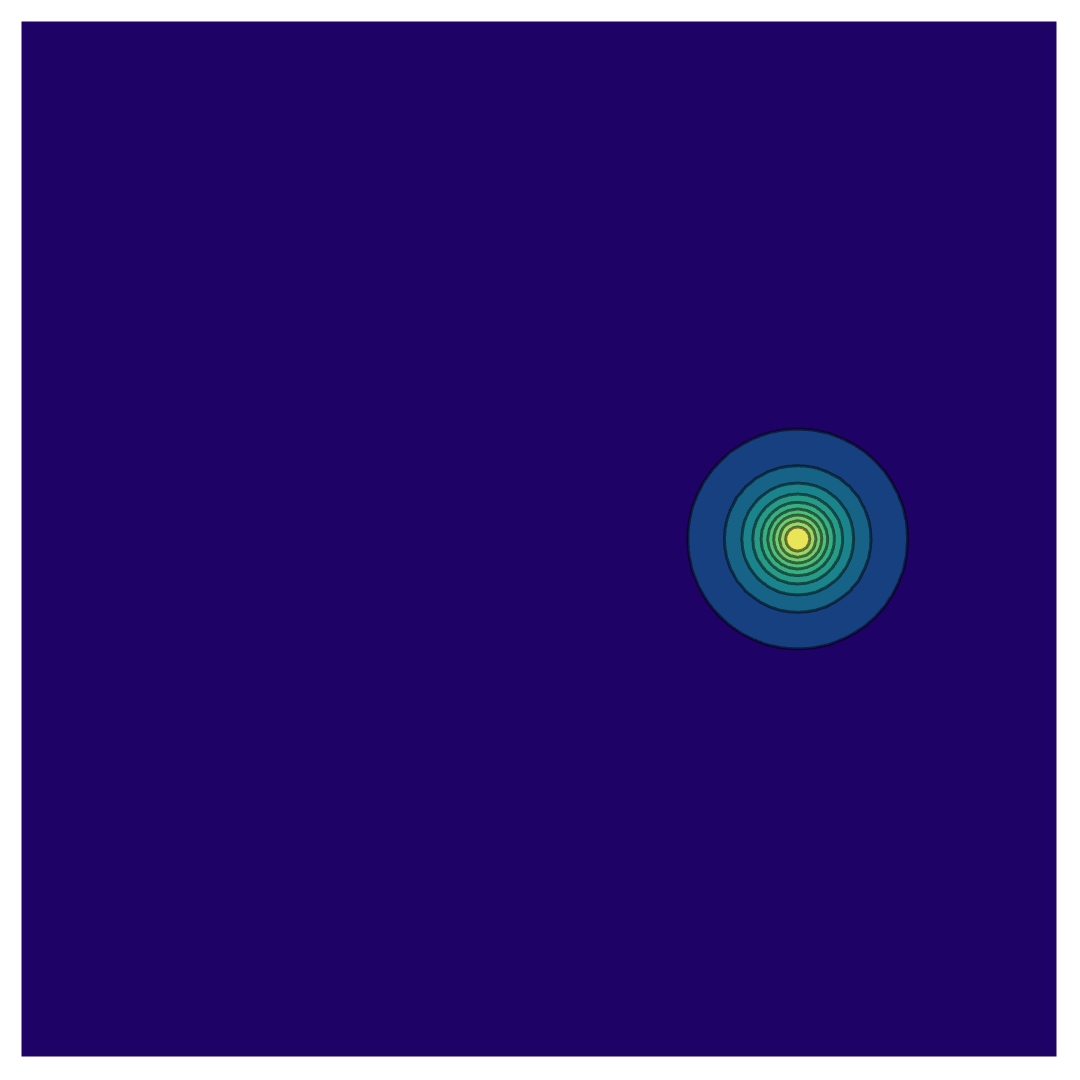}
  \end{subfigure}
  \begin{subfigure}[b]{0.19\textwidth}
    \centering 
    \includegraphics[width=\textwidth]{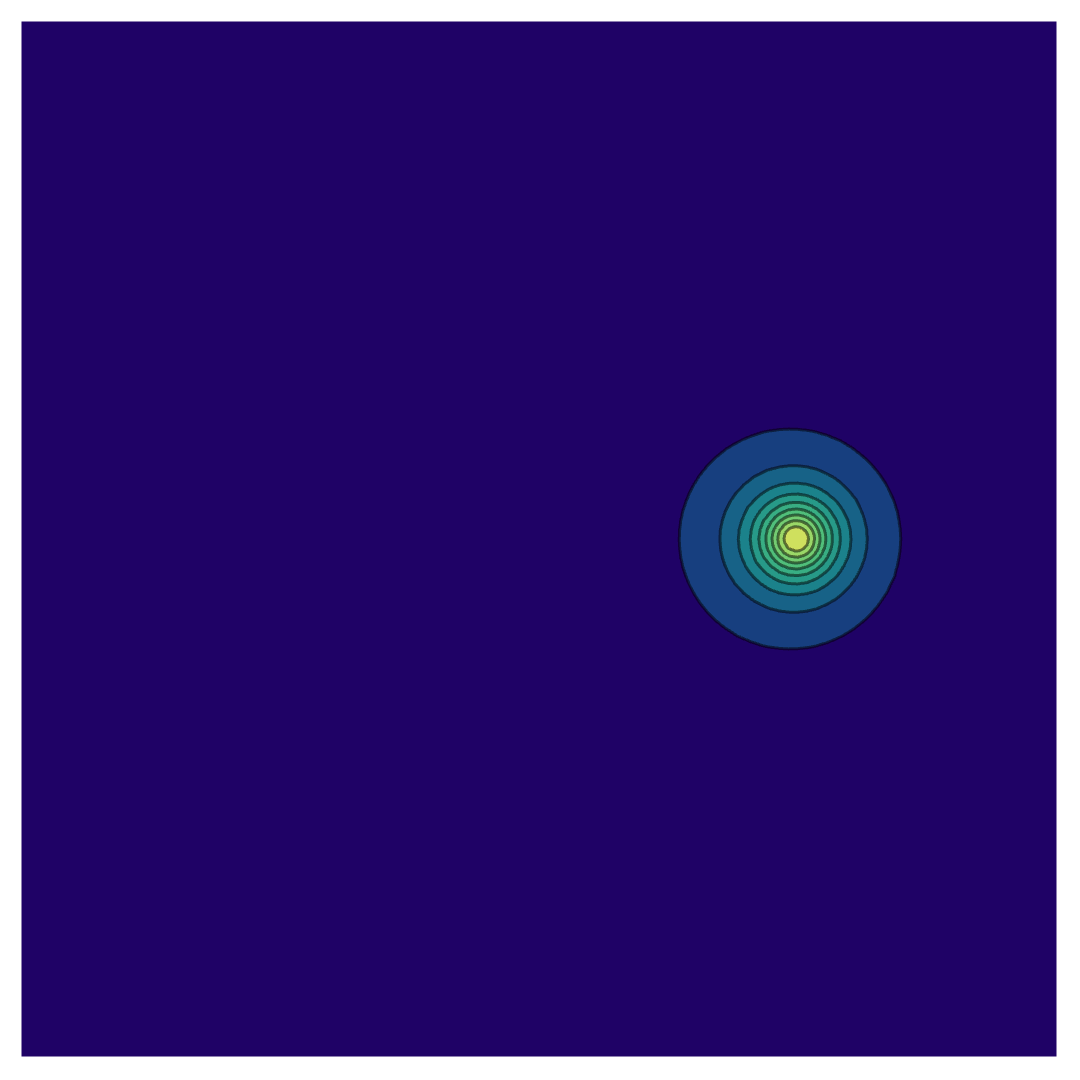}
  \end{subfigure}
  \begin{subfigure}[b]{0.19\textwidth}
    \centering 
    \includegraphics[width=\textwidth]{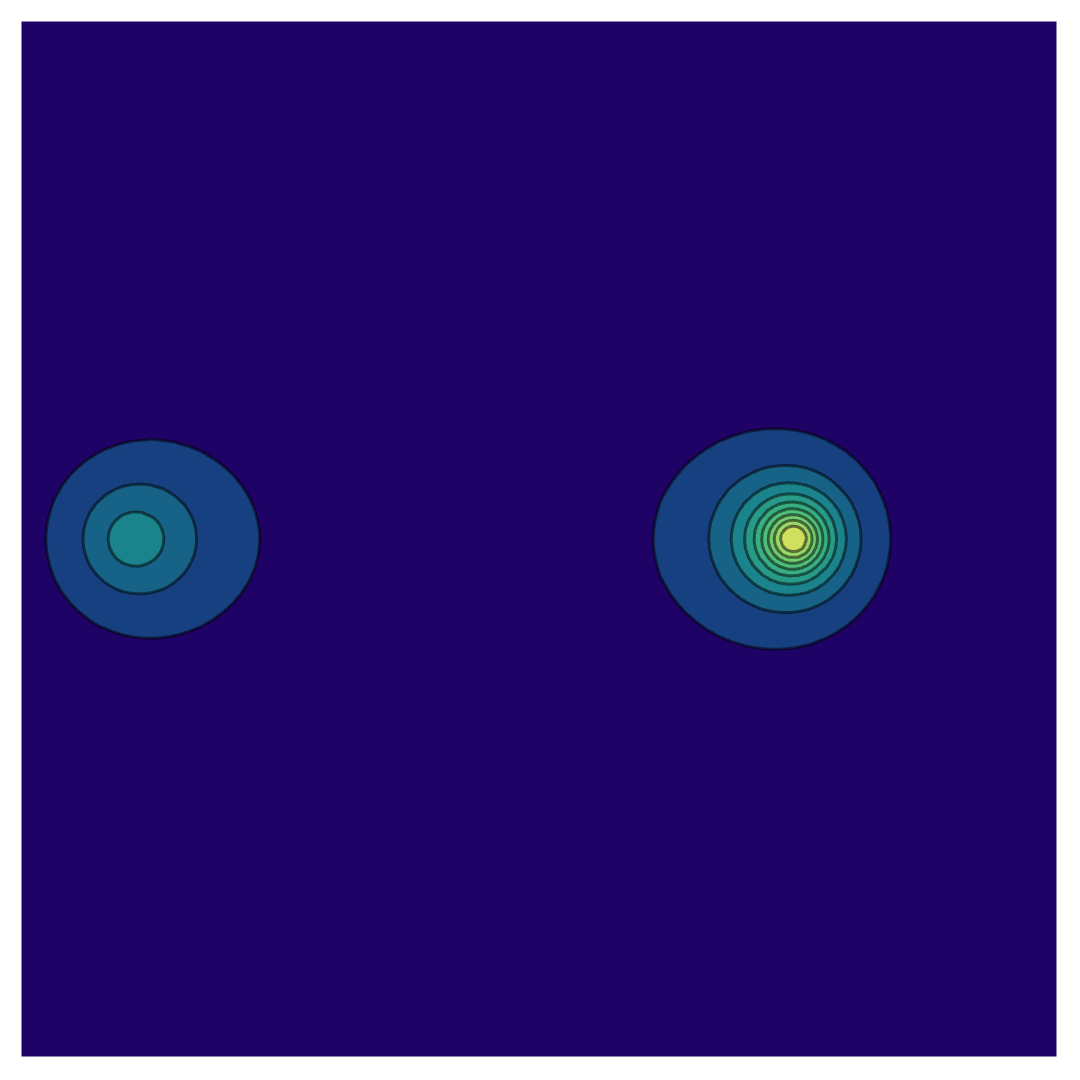}
  \end{subfigure}
  \begin{subfigure}[b]{0.19\textwidth}
    \centering 
    \includegraphics[width=\textwidth]{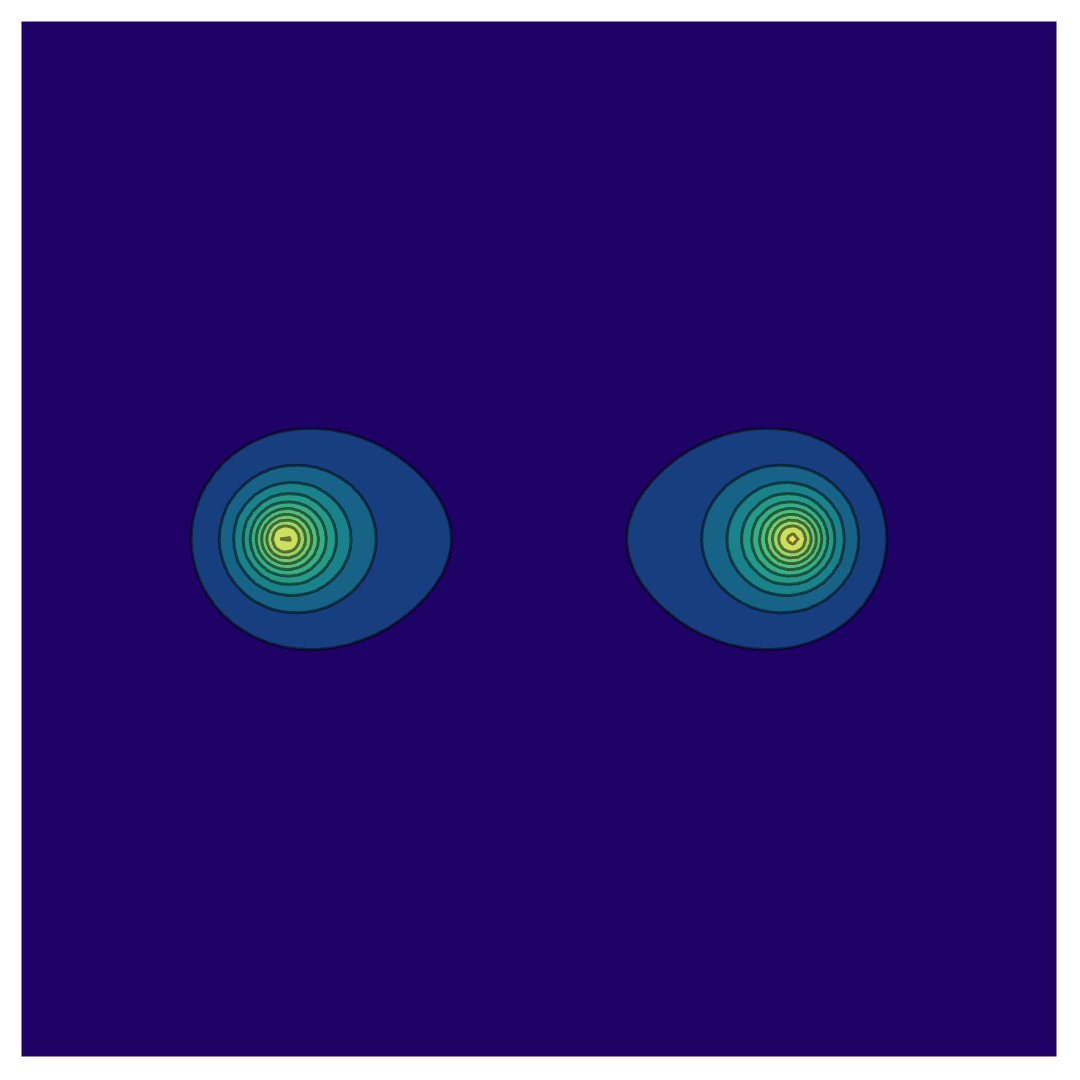}
  \end{subfigure}
  \begin{subfigure}[b]{0.19\textwidth}
    \centering 
    \includegraphics[width=\textwidth]{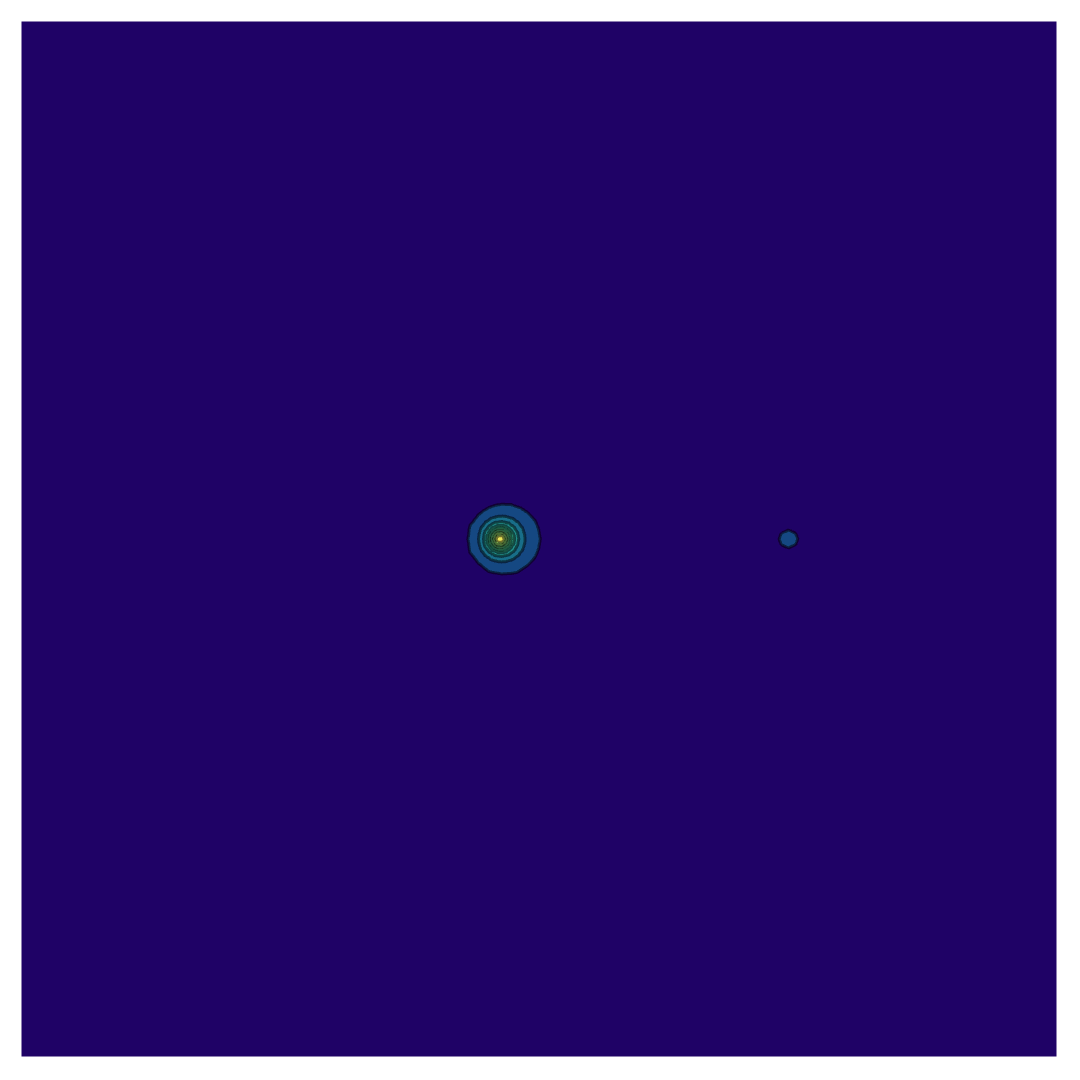}
  \end{subfigure}
  \caption{\small From top to bottom, the value of the deformation constants $k$ are taken to be $5,20,50$, respectively. Also, the value of $b$ in \eqref{c1soln} of the first two rows are chosen to be $0,1/\sqrt{3},1,\sqrt{3},2+\sqrt{3}$ from left to right while the parameters $b$ of the last row are $0,2-\sqrt{3},(\sqrt{2}-1)(\sqrt{3}-\sqrt{2}), 1,\sqrt{3}$, respectively.}
  \label{fig:lumps}
\end{figure*}

Collecting all ingredients above, we can find that the energy density of the baby Skyrmion solution \eqref{c1soln} takes the form
\begin{multline}
  \mathcal{E} = 2\abs{a}^2 \cdot \Big[
    \abs{b(-z_0+x+iy)+a}^4
    \\
    +2k\abs{(b(-z_0+x+iy)+a)(x+iy-z_0)}^2
    +\abs{x+iy-z_0}^4
  \Big]^{-1} \,.
\end{multline}
Some numerical demonstrations are shown in Fig. \ref{fig:lumps}. 
{In Fig. \ref{fig:lumps}, we not only plot the baby Skyrmion configuration for different $k$, but also include the plots of different $b$ with a fixed $k$ since the vacuum moduli are inhomogeneous.}

Summarizing, for a fixed value of $k$, we observe two peaks when the base point in the vacuum moduli is situated at the equator. As we move toward the two poles, these peaks converge into a single peak. Conversely, when we vary the deformation constant $k$, we notice that the two peaks overlap for small values of $k$ (i.e., $k\gsim 1$) and gradually separate into two distinct lumps as $k$ increases. Moreover, the rate of transitioning into a single peak is greater for larger values of $k$.
Note that the separation of the peaks in the current scenario is due to the inhomogeneity of the vacuum moduli, in contrast to the more common occurrence where multiple lumps arise from large topological charges; for a recent example, see \cite{Eto:2023lyo}.

\section{Large-\boldmath{$k$} limit. The sausage model.}
\label{lkl}

In this section we discuss the large-$k$ limit which leads us to the cigar model \cite{FOZ,LZ}. This can be seen by a simple rearrangement of Eq. \eqref{two}, 
\begin{align}
  \mathcal{L} = \frac{1}{g^2\sqrt{k^2-1}} \frac{(\d S_{i})^2}{\sqrt{\frac{k+1}{k-1}} - \sqrt{\frac{k-1}{k+1}}S_3^2}
\end{align}
coinciding with the standard expression of the sausage model in \cite{FOZ,LZ} as long as $k \geq 1$. A sketch of the corresponding metric is presented in Fig. \ref{sausage}. See Appendix \ref{sec:largekplot} for further details.
\begin{figure}[t] 
  \centering 
  \includegraphics[width=.9\linewidth]{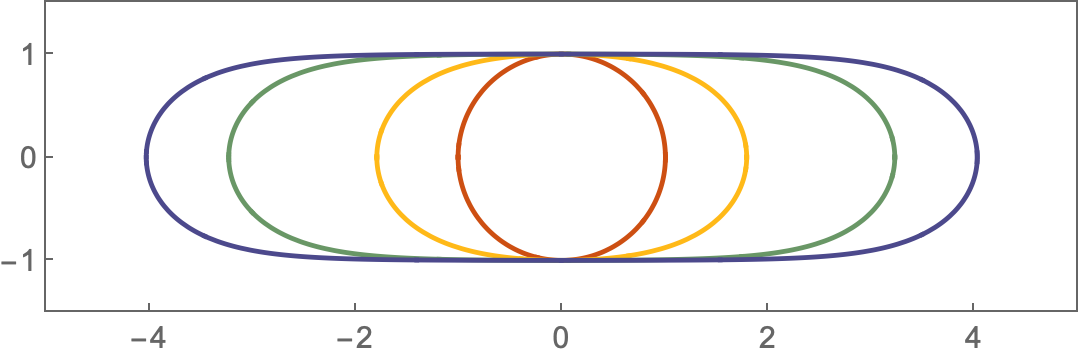}
  \caption{\small The metric of the sausage model in the present parametrization. The contours from inside to outside (i.e. red to blue) correspond to $k$ equal to 1.0, 9.5, 200, and 1000, respectively. The  surfaces corresponding to the metric are embedded in the three-dimensional Euclidean space \cite{Belardinelli:1994dq,FOZ}.}
  See also Appendix \ref{sec:largekplot} for the coordinate $(X,Z)$.
  \label{sausage}
\end{figure}

In the following, we give a short illustration of the target space in the large $k$ limit. As shown in the above illustration, the model is mostly characterized by its length of the flat part. To make an estimation of the length, we rewrite the Lagrangian of the deformed CP(1) model as follows:
\begin{align}\label{Lsph}
  \mathcal{L} = \frac{1}{g^2(1+k)}\frac{(\d\alpha)^2 + \sin^2{\alpha}(\d\beta)^2}{1 - \frac{k-1}{k+1}\cos^{2}{\alpha}}
\end{align}
As $k \gg 1$, we want to compare \eqref{Lsph} with the metric of a cylindrical shell (up to an overall constant)
\begin{align}
  \dd{l}^2 + a^2 \dd{\beta}^2
\end{align}
where $l$ is the coordinate along the cylinder and $a$ is the radius of the shell. To proceed, let us make a change of the variable of \eqref{Lsph} such that for the deformed CP(1)
\begin{align}
  \cL \sim \dd{\zeta}^2 + r(\zeta)^2 \dd{\beta}^2 \,.
\end{align}
Both $\zeta$ and $r(\zeta)$ are of dimension $[m^0]$.
In the following we will see that the radius $r(\zeta) \sim 1$ for a certain range of the coordinate along the cylinder $\zeta$ which we can take twice of this range as the ``length'' of the cylinder. 
To see this is the case, we note that 
\begin{align}
  \zeta = -\int_{\frac{\pi}{2}}^{\frac{\pi}{2}-\alpha} \frac{\dd{t}}{\sqrt{1-\frac{k-1}{k+1}\cos^2{t}}}
  = - F\left( \frac{\pi}{2}-\alpha,\,\frac{k-1}{k+1} \right)
\end{align}
where $F(\theta,\kappa)$ is the incomplete elliptic integral defined as 
\begin{align}\label{inF}
  F(\theta,\kappa) := \int_{0}^{\theta} \frac{\dd{t}}{\sqrt{1-\kappa\cos^{2}{t}}} \,.
\end{align}
In addition, the new parameter $\zeta$ varies in the ranges
\begin{align}
  -K\left( \frac{k-1}{k+1} \right) \,\leq\, \zeta \,\leq\, K\left( \frac{k-1}{k+1} \right)
\end{align}
where $K(\kappa)$ is the complete elliptic integral, i.e., $\theta=\pi/2$ in \eqref{inF}. Then $g_{\beta\beta}$ can be expressed in terms of 
\begin{align}
  \frac{\sin^2{\beta}}{1 - \frac{k-1}{k+1}\cos^2{\beta}} 
    =
    \left[ \frac{\cn{\zeta,\frac{k-1}{k+1}}{}}{\dn{\zeta,\frac{k-1}{k+1}}{}}\right]^2
\end{align}
Note that some Jacobi elliptic functions are used in the above reparametrization. Their definitions are as follows. Namely, 
\begin{align}
  \sn{z,\kappa}{} &:= \sin(F^{-1}(z,\kappa))
  \notag\\
  \cn{z,\kappa}{} &:= \cos(F^{-1}(z,\kappa))
\end{align}
with the identity
\begin{align}
  \kappa\sn{z,\kappa}{2}+\dn{z,\kappa}{2} = 1 \,
\end{align}
where $F^{-1}(z,\kappa)$ is the inverse function for the incomplete elliptic function $F$. Now, by direction comparison, we can identify $r(\zeta)$ as 
\begin{align}
  r(\zeta) = \frac{\cn{\zeta,\frac{k-1}{k+1}}{}}{\dn{\zeta,\frac{k-1}{k+1}}{}}
\end{align}
From Fig. \ref{sausage}, we can see that $r(\zeta) \sim 1$ for the cylindrical part, and its derivative has the asymptotic form

\begin{align}
  r'(\zeta) = -\frac{\sinh{2\zeta}}{k} + \order{k^{-2}}
\end{align}
in the large $k$ limit. Note that for any values of $\zeta$ not scaling with $k$, the radius of the cross-sectional circle remains one. 
The expression  $k^{-1} \sinh{2\zeta} $
 becomes $\order{1}$ as $\zeta$ reaches the edge of the cylinder. Therefore, we can then define the half-length of the cylinder by finding the domain 
in which  $\sinh{2\zeta}$ is comparable to $k$, i.e.,
\begin{align}\label{eq:lapprox}
  \mathrm{length} \sim \arcsinh{k}
  \sim \log{k} \,.
\end{align}

\section{Renormalization group flow in 2D}
\label{rentwod}

After examining (quasi)classical aspects of the deformed CP(1) model, particularly its static solutions in 2+1 dimensions in the previous sections, we will now study the role of quantum fluctuations at one-loop level in {\em two} dimensions. One can consider either Euclidean or Minkowski spacetime. We will limit ourselves to $D= 1+1$. In this case loop corrections are logarithmic. 

Strictly speaking, the theory defined by Eqs. (\ref{1one}) and (\ref{1two}) is not renormalizable in the usual sense of this word because the metric 
(\ref{1one}) is not of the Einstein type. However, $G_{1\bar 1} $ in the bare Lagrangian at one loop is replaced by 
\beq
 {\mathcal L}_{\rm eff} =\left\{  G_{1\bar 1} +\frac{1}{2\pi}  G_{1\bar 1}\, {\textstyle\frac{1}{  2}} {\mathcal  R}\, L\right\}\pt_\mu\bar\phi\pt^\mu\phi
 \label{43}
\eeq
where $\Rc$ is the scalar curvature,
\beq
{\mathcal R} =2 G_{1\bar 1}\left(n_1n_2 +4n_1n_3\phi\bar\phi + n_2n_3 \phi^2\bar\phi^2
\right)
\label{44}
\eeq
and 
\beq
L=-\log\frac{M_{\rm uv}}{ \mu}
\eeq
is the renormalization group (RG) ``time."
Equations (\ref{43}) and (\ref{44}) follow from the general theory of nonlinear sigma models \cite{SK}, which implies that the $\beta$ function of the metric at the one-loop level is proportional to its Ricci tensor. Assembling together Eqs. (\ref{1one}), (\ref{twop}), (\ref{43}), and (\ref{44}) we obtain
 \begin{align}
  \delta G_{1\bar{1}} 
  &=- \left(G_{1\bar{1}}\right)^2\, \delta  G^{\bar{1}1} 
  \nonumber\\
  &=\left(n_1n_2 +4n_1n_3\phi\bar\phi + n_2n_3 \phi^2\bar\phi^2
  \right)
  \left(- \frac{1}{2\pi} L\right)  \left(G_{1\bar{1}}\right)^2
 \end{align}
implying 
\begin{align}
  \delta  G^{\bar{1}1}= \left(n_1n_2 +4n_1^2\phi\bar\phi + n_1n_2 \phi^2\bar\phi^2 \right)\frac{1}{2\pi}L\,.
  \label{46}
 \end{align}
 This equation implies
 \begin{align}\label{47}
  \dv{g^2}{L} = -\frac{kg^4}{2\pi} \,,\quad
  \dv{(g^2k)}{L} = -\frac{g^4}{2\pi}\,,
\end{align}
following from Eq. \eqref{1one}. 
In passing from (\ref{46}) to (\ref{47}) we compared the right-hand side of (\ref{46}) with the coefficients in $G^{\bar{1}1} =1 /G_{1\bar 1}$. Also, we assume $n_{1,3}$ are nonsingular and apply the identification \eqref{twop}.

The RG flow equations for $g^2$ and $k$ are summarized by the following $\beta$ function:
\beq
\beta_{G_{1\bar 1}} = G_{1\bar 1}\, \frac{{\textstyle\frac{1}{ 2}} {\mathcal R}}{2\pi} \,\frac{1}{1-\frac{1}{2\pi} {\textstyle\frac{1}{ 2}}{\mathcal R}}\,.
\label{48}
\eeq
In fact, in this paper we derived only the one-loop expression corresponding to the term $O(\Rc )$ in the expansion of the 
right-hand side in (\ref{48}). As a matter of fact, if we continue expansion up to $O(\Rc^2 )$ we will correctly reproduce the two-loop result, too.

\begin{figure}[t] 
  \centering 
  \includegraphics[width=.6\linewidth]{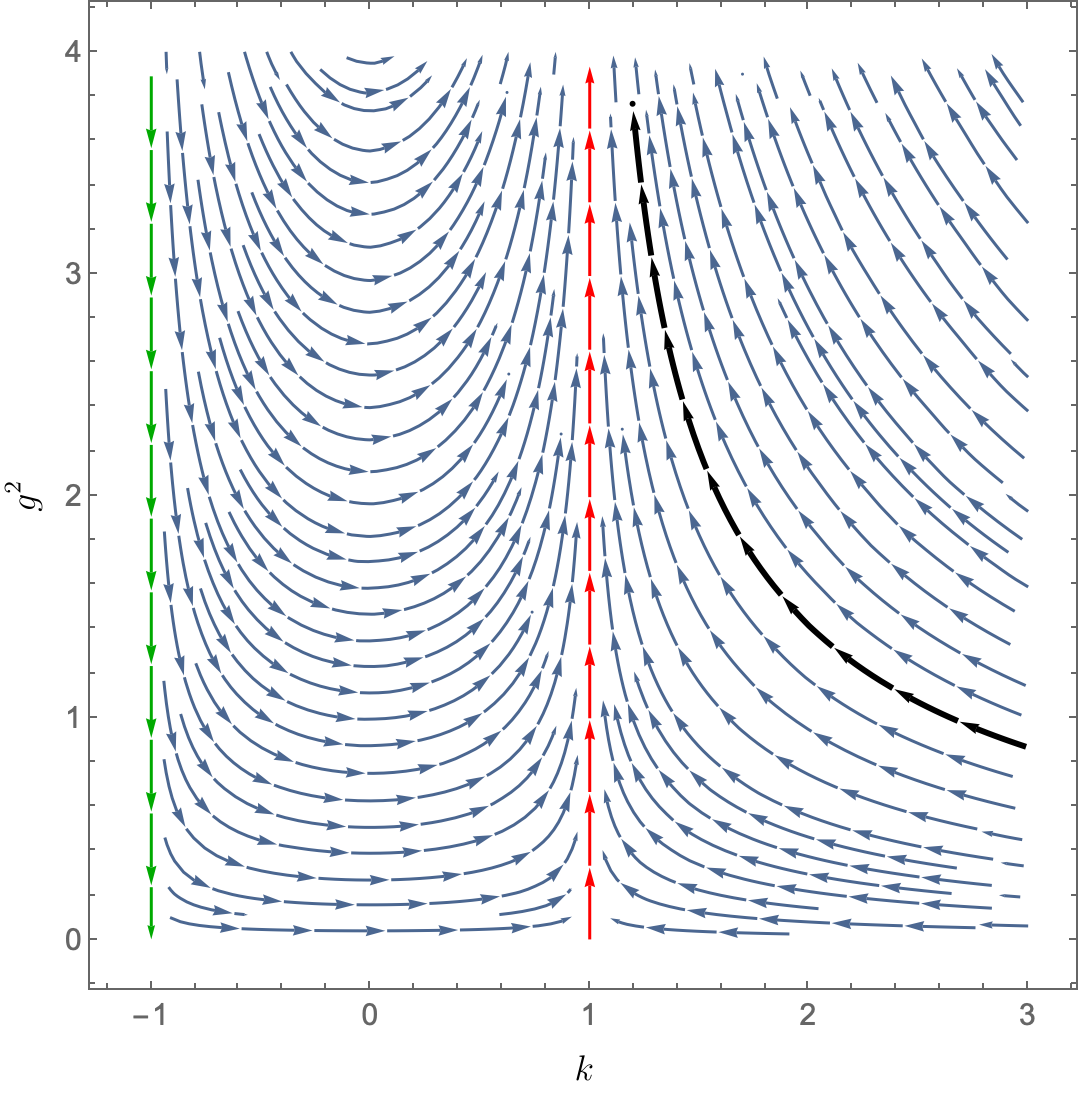}
  \caption{\small The RG flow of $g^2$ and $k$ couplings. The flow is from UV to IR. The separatrices are marked with green and red lines while the thick black line represents an example of the RG flow in the limit of the  sausage model  (i.e., large $k$).}
  \label{fig:rgflow}
\end{figure}

The two coupling constants $g^2$ and $k$ are coupled 
under the running flow. The combination
\beq
g^4(k^2-1)= {\rm RGI}
\eeq
where RGI stands for RG invariant. We integrate (\ref{47}) numerically to obtain the RG flow from UV to IR. The result of integration  is shown 
in Fig. \ref{fig:rgflow}. The directions of the arrows indicate the flow to IR. $k=1$ is a separatrix. On the other hand, the other separatrix shows up at $k=-1$, i.e., the green one in Fig. \ref{fig:rgflow}. This can be seen from the RG flow equation of $k$, namely,
\begin{align}
  \dv{k}{L} \sim k^2-1 \,.  
\end{align}
Qualitatively, the RG flows can be categorized into two groups. On the right-hand side of the $k=1$ separatrix (in the limit of large $k$), the initial state with a small $g^2$ evolves into a state with strong coupling and $k=1$. See, for example, the thick black line in Fig. \ref{fig:rgflow}. On the other hand, when the UV starting point of the RG flow formula lies between the two separatrices ($-1 \leq k \leq 1$), the coupling weakens initially and eventually becomes strong as the elongation parameter approaches unity.
Thus in the IR we return to the CP(1) model and the full O(3) symmetry is restored.

\section{Conclusions and perspectives}
\label{concl}

In this paper we consider some implications of the Lie-algebraic deformations of the CP(1) model (see \cite{Los}). First, we present a one-parametric family 
of deformations --the simplest example of the Lie-algebraic models; see Eqs. (\ref{two}) and (\ref{three}). In addition to the coupling constant $g^2$ of CP(1), 
an ``elongation" parameter $k$ is added.
In this example, the O(3) symmetry of the CP(1) model is broken down to O(2) by construction. The target space is K\"ahlerian; we find the corresponding K\"ahler function (\ref{eq:kahler}). This class of models, being considered in $D=1+1$
supports deformed Polakov-Belavin instantons which in $D=1+2$ (in the static limit) become baby Skyrmions. We determine the instanton action and baby Skyrmion mass as a function of $g^2$ and $k$. In Sec. 4 the internal structure of the
deformed PB instantons is studied; see Figs. 2 and 3.

Finally, we explore in detail two limiting cases:
\beq
k\gsim 1\,,\quad {\rm and}\quad k\gg 1\,.
\eeq
In the first case, as we approach $k=1$ we recover the CP(1) model. In the second limit we approach the so-called cigar or sausage model, see Sec. 5.
The interpolation between these two extreme cases is smooth.

In $D=2$ (in which case renormalizations are logarithmic) we calculate the one-loop ``$\beta$ function."   The quotation marks emphasize the fact that  the $\beta$ function is rather peculiar (see Sec. 6). The geometric structure
of the one-loop Lagrangian is different from that of the classical Lagrangian. Nevertheless, no new coupling constants appear --the constants $g^2$ and $k$ are entangled in the running formula. Moreover, in the infrared, $g^2$ runs toward the strong coupling domain, while $k$ approaches 1; see Fig, 6.

In conclusion, let us briefly discuss possible phenomenological uses of the Lie-algebraic sigma models --the simplest nontrivial example presented in Eqs. (\ref{two}) and (\ref{three}), or more general versions, with more than three parameters introduced in \eqref{1one}. So far, we are aware of a single appropriate example of magnetoelectric effects in Mott insulators. For simplicity, let us limit our consideration to the window $|k-1|\ll1$.  Then Eqs. (\ref{two}) and (\ref{three}) can be represented as 
\beqn
\label{51}
\Lc &=&\frac{1}{2\tilde{g}^2} (\pt S_i) (\pt S_i)\left\{1+\varepsilon S_3^2
\right\}
\,,\qquad \vec{S} \vec{S}=1\,.
\\[1mm]
\tilde{g}^2 &=& g^2\frac{k+1}{2}\,, \quad\varepsilon = \frac{k-1}{2}\,,
\nonumber
\eeqn
up to terms of higher order in the small parameter $\varepsilon$. The first term in the first line in \eqref{51} is isotropic in the target space [i.e. SU(2) invariant]. Switching on the $\varepsilon$ term breaks the SU(2) symmetry of the target space down to U(1). Unisotropic deformations which are currently studied in experiments (see, e.g., \cite{pand}) are expected to pave the way to new exotic magnetic states. For instance, Ref. \cite{zlb}  deals with a dynamical material system  which can reduce to \eqref{51} in the continuum limit (under certain fine-tuning of parameters).

The above-mentioned system of strongly coupled Mott insulators includes spin-orbit couplings that produce an impact on electronically induced magnetoelectric effects. In addition to the isotropic Heisenberg term [see Eq. (11) in \cite{zlb}] biquadratic interactions appearing in two-orbital versions
lead to an extra quartic term of the following form  \cite{zlb},
\beqn
{\mathcal H}_{\rm quart} &\propto& \sum_{(k,j)\rm neighbors} \Big[S^{(k)}_3S^{(j)}_3 +\Delta \Big(S^{(k)}_1S^{(j)}_1+S^{(k)}_2 S^{(j)}_2
\Big)
\Big]\nonumber\\[2mm]
&\times&  \Big[S^{(k)}_3S^{(j)}_3 +\Delta^\prime \Big(S^{(k)}_1S^{(j)}_1+S^{(k)}_2 S^{(j)}_2
\Big)
\Big]
\eeqn
on the square 2D lattice. The upper indices in the  parentheses mark the nodes of this lattice. If $\Delta\to 1$ and $\Delta^\prime \to 0$ in the continuum limit this  returns us to the model \eqref{51}.

As far as the opposite limit $k\gg1$ is concerned, we hope 
that the emerging cigarlike baby Skyrmions/PB instantons could be used in exotic magnetic phenomena in condensed matter and in high-energy theory,
but so far this is not yet clear.

\section*{Acknowledgements}

We are grateful to C. Batista, O. Gamayun, A. Losev, V. Lukyanov,  M. Nitta, N. Perkins and O. Starykh  for very useful communications.

This work is supported in part by DOE Grant No. DE-SC0011842 and by Fulbright Scholarship. CS is supported by Larkin Fellowship.
This work was completed during MS visit at 
the Institute for Particle and Nuclear Physics, Charles University (Prague, Czech Republic). MS thanks his colleagues from Charles University
for kind hospitality.

\appendix 
\section{Embedding of the target manifold in large $k$ limit}
\label{sec:largekplot}

In this appendix, we estimate the length of the flat part of the target manifold. In this section, we detail the process to find the surface evolution.

Along the same line with \cite{Belardinelli:1994dq}, we embed the target manifold in a three-dimensional Euclidean space $\RR^3$ such that the induced metric coincides with \eqref{Lsph}. Assume that $(X,Y,Z)$ is the coordinate of $\RR^3$ and $Z$ is the axis of symmetry. It then suffices to consider the section of $X$ and $Z$ because of the $U(1)$ symmetry. To parametrize $X$ and $Z$, we can first consider $\alpha \equiv {\rm const}$, resulting in 
\begin{align}
  X = \frac{\pm\sin{\alpha}}{\sqrt{1-\frac{k-1}{k+1}\cos^2{\alpha}}}
  \,.
\end{align}
On the other hand, as $\beta \equiv {\rm const}$, we can find $Z$ via 
\begin{align}
  \dd{Z}^2 = \left[  
    \frac{1}{1-\frac{k-1}{k+1}\cos^2{\alpha}}
    - \left( \dv{X}{\alpha} \right)^2
    \right] \dd{\alpha}^2
\end{align}
which turns out to be 
\begin{align}
  Z = 
  i\sqrt{\frac{k+1}{k-1}} 
  E\left( i\gamma \,\Big|\, \frac{2}{k+1}  \right)
\end{align}
where 
\begin{align*}
  \csch\gamma = \sqrt{\frac{k+\cosh\left( 2\log{\tan(\alpha/2)} \right)}{k-1}} 
\end{align*}
and $E(a|b)$ is the elliptic integral of the second kind.
With $(X,Z)$ parametrized in $\alpha$, one can find the illustration of the target space in Fig. \ref{sausage} in which $\alpha$ ranges from 0 to $\pi$.

Furthermore, to approximate the length of the cylindrical part of the target manifold, one observes that for large enough $k$, $\abs{X}$ drops rapidly at both ends. Therefore, the length scales as 
\begin{align}
  Z(\alpha=\pi) - Z(\alpha=0)
  \sim \log{2k}
\end{align}
consistent with what we have found in \eqref{eq:lapprox}.

\end{document}